\documentclass[prd,article,nofootinbib,twocolumn,preprintnumbers,superscriptaddress,amsmath,amssymb,aps]{revtex4-1}

\usepackage{graphicx,
longtable,
color,
multirow
}

\begin{document}

\preprint{IPPP/20/35}

\title{Non-standard neutrino interactions as a solution to the NO$\nu$A and T2K discrepancy}%

\author{		Sabya Sachi Chatterjee}
\email{sabya.s.chatterjee@durham.ac.uk}
\affiliation{		Institute for Particle Physics Phenomenology, Department of Physics, Durham University, Durham, DH1 3LE, UK}

\author{		Antonio Palazzo}
\email{palazzo@ba.infn.it}
\affiliation{ 	Dipartimento Interateneo di Fisica ``Michelangelo Merlin,'' Via Amendola 173, 70126 Bari, Italy}
\affiliation{ 	Istituto Nazionale di Fisica Nucleare, Sezione di Bari, Via Orabona 4, 70126 Bari, Italy}


\begin{abstract}

The latest data of the two long-baseline accelerator experiments NO$\nu$A and T2K, interpreted in the standard 3-flavor scenario, display a discrepancy. A mismatch in the determination of the standard CP-phase $\delta_{\mathrm {CP}}$ extracted by the two experiments is evident in the normal neutrino mass ordering. While NO$\nu$A prefers values close to $\delta_{\mathrm {CP}} \sim 0.8 \pi$, T2K identifies values of $\delta_{\mathrm {CP}} \sim 1.4 \pi$. Such two estimates are in disagreement at more than 90$\%$ C.L. for 2 degrees of freedom. We show that such a tension can be resolved if one hypothesizes the existence of complex neutral-current non-standard interactions (NSI) of the flavor changing type involving the $e-\mu$ or the $e-\tau$ sectors with couplings $|\varepsilon_{e\mu}| \sim |\varepsilon_{e\tau}|\sim
0.2$. Remarkably, in the presence of such NSI, both experiments point towards the same 
common value of the standard CP-phase $\delta_{\mathrm {CP}} \sim 3\pi/2$. Our analysis also highlights an intriguing preference for maximal CP-violation in the non-standard sector with the  NSI CP-phases having best fit close to $\phi_{e\mu}  \sim \phi_{e\tau}\sim 3\pi/2$,
hence pointing towards imaginary NSI couplings. 
   
\end{abstract}
\maketitle

{\bf {\em Introduction.}}  The two long-baseline (LBL) accelerator experiments NO$\nu$A and T2K have recently released new
data at the Neutrino 2020 Conference~\cite{NOVA_talk_nu2020,T2K_talk_nu2020}. Intriguingly, the 
two experiments display a moderate tension
preferring values of the standard 3-flavor CP-phase $\delta_{\mathrm {CP}}$ which are in disagreement. 
While this discrepancy may be imputable to a statistical fluctuation or to an unknown systematic error, it
may represent an indication of physics beyond the Standard Model (SM). In particular, one should note
that the two experiments are different with respect to their sensitivity to the matter effects, due to the
different baselines (810 km for NO$\nu$A and 295 km for T2K). This evokes the fascinating possibility that
new physics may be at work in the form of non-standard neutrino interactions (NSI).

{\bf {\em Theoretical framework.}} NSI may constitute the low-energy manifestation of high-energy physics of new
heavy states (for a review see~\cite{Farzan:2017xzy,Biggio:2009nt,Ohlsson:2012kf,Miranda:2015dra,Dev:2019anc}) or, 
they can be related to light mediators~\cite{Farzan:2015doa,Farzan:2015hkd}.
As first noted in~\cite{Wolfenstein:1977ue}, NSI can alter the dynamics~\cite{Wolfenstein:1977ue,Mikheev:1986gs,Mikheev:1986wj} of the neutrino flavor conversion in matter. 
The presence of NSI can have a sizeable impact on the interpretation of current LBL data. Notably, in the recent work~\cite{Capozzi:2019iqn}, it has been evidenced that they may even obscure the correct determination of the neutrino mass ordering (NMO).%
\footnote{In the 3-flavor scheme there are three mass eigenstates $\nu_i$ with masses
$m_i\, (i = 1,2,3)$, three mixing angles $\theta_{12},\theta_{13}, \theta_{13}$, and one CP-phase $\delta_{\mathrm {CP}}$.
The mass ordering is defined to be normal (inverted) if $m_3>m_{1,2}$  ($m_3<m_{1,2}$).
We will abbreviate normal (inverted) ordering as NO (IO).}
The impact of NSI on present and future new-generation LBL experiments has been widely explored (see for example~\cite{Friedland:2012tq,Coelho:2012bp,Girardi:2014kca,Rahman:2015vqa,Coloma:2015kiu,deGouvea:2015ndi,Agarwalla:2016fkh,Liao:2016hsa,Forero:2016cmb,Huitu:2016bmb,Bakhti:2016prn,Masud:2016bvp,Soumya:2016enw,Masud:2016gcl,deGouvea:2016pom,Fukasawa:2016lew,Liao:2016orc,Liao:2016bgf,Blennow:2016etl,Deepthi:2017gxg,Flores:2018kwk,Hyde:2018tqt,Masud:2018pig,Esteban:2019lfo}.)
The NSI can be represented by a dimension-six operator~\cite{Wolfenstein:1977ue}
\begin{equation}
\mathcal{L}_{\mathrm{NC-NSI}} \;=\;
-2\sqrt{2}G_F 
\varepsilon_{\alpha\beta}^{fC}
\bigl(\overline{\nu_\alpha}\gamma^\mu P_L \nu_\beta\bigr)
\bigl(\overline{f}\gamma_\mu P_C f\bigr)
\;,
\label{H_NC-NSI}
\end{equation}
where $\alpha, \beta = e,\mu,\tau$ indicate the 
neutrino flavor,  $f = e,u,d$ denote the matter 
fermions, $P$ represents the projector operator with
superscript $C=L, R$ referring to the chirality of the 
$ff$ current, and $\varepsilon_{\alpha\beta}^{fC}$ are the strengths 
of the NSI. The hermiticity of the interaction implies
\begin{equation}
\varepsilon_{\beta\alpha}^{fC} \;=\; (\varepsilon_{\alpha\beta}^{fC})^*
\;.
\end{equation}
For neutrino propagation in the Earth, the relevant combinations are
\begin{equation}
\varepsilon_{\alpha\beta}
\;\equiv\; 
\sum_{f=e,u,d}
\varepsilon_{\alpha\beta}^{f}
\dfrac{N_f}{N_e}
\;\equiv\;
\sum_{f=e,u,d}
\left(
\varepsilon_{\alpha\beta}^{fL}+
\varepsilon_{\alpha\beta}^{fR}
\right)\dfrac{N_f}{N_e}
\;,
\label{epsilondef}
\end{equation}
$N_f$ being the number density of $f$ fermion.
For the Earth, we can consider neutral and isoscalar matter, with  $N_n \simeq N_p = N_e$, 
in which case $N_u \simeq N_d \simeq 3N_e$.
Therefore,
\begin{equation}
\varepsilon_{\alpha\beta}\, \simeq\,
\varepsilon_{\alpha\beta}^{e}
+3\,\varepsilon_{\alpha\beta}^{u}
+3\,\varepsilon_{\alpha\beta}^{d}
\;.
\label{epsilon_eff}
\end{equation}
The NSI alter the effective Hamiltonian of neutrino propagation 
in matter, which in the flavor basis reads
\begin{equation}
H \;=\; 
U
\begin{bmatrix} 
0 & 0 & 0 \\ 
0 & k_{21}  & 0 \\ 
0 & 0 & k_{31} 
\end{bmatrix}
U^\dagger
+
V_{\mathrm{CC}}
\begin{bmatrix}
1 + \varepsilon_{ee}  & \varepsilon_{e\mu}      & \varepsilon_{e\tau}   \\
\varepsilon_{e\mu}^*  & \varepsilon_{\mu\mu}    & \varepsilon_{\mu\tau} \\
\varepsilon_{e\tau}^* & \varepsilon_{\mu\tau}^* & \varepsilon_{\tau\tau}
\end{bmatrix}\,,
\end{equation}
where $U$ is the Pontecorvo-Maki-Nakagawa-Sakata (PMNS) matrix, which
depends on three mixing angles ($\theta_{12}, \theta_{13}, \theta_{23}$) and the CP-phase $\delta_{\mathrm {CP}}$.
The parameters $k_{21} \equiv \Delta m^2_{21}/2E$ and $k_{31} \equiv \Delta m^2_{31}/2E$ represent
the solar and atmospheric wavenumbers, where $\Delta m^2_{ij} \equiv m^2_i-m^2_j$, while
$V_{\mathrm{CC}}$ is the charged-current matter potential 
\begin{equation}
V_{\mathrm{CC}} 
\;=\; \sqrt{2}G_F N_e 
\;\simeq\; 7.6\, Y_e \times 10^{-14}
\bigg[\dfrac{\rho}{\mathrm{g/cm^3}}\bigg]\,\mathrm{eV}\,,
\label{matter-V}
\end{equation}
where $Y_e = N_e/(N_p+N_n) \simeq 0.5$ is the relative electron number density in the Earth crust.
It is useful to introduce the dimensionless quantity $v = V_{\mathrm{CC}}/k_{31}$, which 
measures the sensitivity to matter effects. Its absolute value
\begin{equation}
|v| 
\;=\; \bigg|\frac{V_{\mathrm{CC}}}{k_{31}}\bigg| 
\;\simeq\; 8.8 \times 10^{-2} \bigg[\frac{E}{\mathrm{GeV}}\bigg]\;,
\label{matter-v}
\end{equation}
will appear in the expressions of the $\nu_\mu \to \nu_e$ conversion probability.
We here emphasize that in T2K (NO$\nu$A) the first oscillation maximum is reached respectively 
for $E \simeq 0.6\, {\mathrm{GeV}}$  ($E \simeq1.6\, {\mathrm{GeV}}$).  This implies that matter effects
are a factor of three bigger in NO$\nu$A ($v  \simeq0.14$) than in T2K ($v  \simeq 0.05$). This
suggests that NO$\nu$A may be sensitive to NSI to which T2K is basically insensitive, so
explaining the apparent disagreement among the two experiments when their results are interpreted in the 
standard 3-flavor scheme.

In the present manuscript, we focus on flavor non-diagonal NSI, that is
$\varepsilon_{\alpha\beta}$'s with $\alpha\ne\beta$. We remark that only such
flavor-changing NSI carry out a dependency on a new CP-phase, which is a crucial ingredient
to resolve the discrepancy between NO$\nu$A and T2K we are considering. 
More specifically, we consider the couplings $\varepsilon_{e\mu}$
and $\varepsilon_{e\tau}$, which, as will we discuss below, introduce a dependency 
on their associated CP-phase in the appearance $\nu_\mu \to \nu_e$ probability%
\footnote{The $\nu_\mu \to \nu_\mu$ disappearance channel is sensitive to the ${\mu-\tau}$ NSI but 
this can be safely ignored because of the very strong upper bound
 put by the atmospheric neutrinos $|\varepsilon_{\mu\tau}| < 8.0 \times 10^{-3}$~\cite{Aartsen:2017xtt}
(see also \cite{Mitsuka:2011ty}).}.
 Let us focus on the conversion probability relevant for the LBL experiments T2K and NO$\nu$A.
In the presence of NSI, the probability can be  expressed as 
the sum of three terms~\cite{Kikuchi:2008vq} 
\begin{eqnarray}
\label{eq:Pme_4nu_3_terms}
P_{\mu e}  \simeq  P_{\rm{0}} + P_{\rm {1}}+   P_{\rm {2}}\,,
\end{eqnarray}
which, using a compact notation similar to~\cite{Liao:2016hsa}, take the following forms
\begin{eqnarray}
\label{eq:P0}
 & P_{\rm {0}} &\,\, \simeq\,  4 s_{13}^2 s^2_{23}  f^2\,,\\
\label{eq:P1}
 & P_{\rm {1}} &\,\,  \simeq\,   8 s_{13} s_{12} c_{12} s_{23} c_{23} \alpha f g \cos({\Delta + \delta_{\mathrm {CP}}})\,,\\
 \label{eq:P2}
 & P_{\rm {2}} &\,\,  \simeq\,  8 s_{13} s_{23} v |\varepsilon|   
 [a f^2 \cos(\delta_{\mathrm {CP}} + \phi) + b f g\cos(\Delta + \delta_{\mathrm {CP}} + \phi)]\,,\nonumber\\
\end{eqnarray}
where $\Delta \equiv  \Delta m^2_{31}L/4E$ is the atmospheric oscillating frequency,
$L$ is the baseline and $E$ the neutrino energy, and $\alpha \equiv \Delta m^2_{21}/ \Delta m^2_{31}$.
For brevity, we have used the notation ($s_{ij} \equiv \sin \theta_{ij} $, $c_{ij} \equiv \cos \theta_{ij}$), 
and following~\cite{Barger:2001yr},
we have introduced 
\begin{eqnarray}
\label{eq:S}
f \equiv \frac{\sin [(1-v) \Delta]}{1-v}\,, \qquad  g \equiv \frac{\sin v\Delta}{v}\,.
\end{eqnarray}
In Eq.~(\ref{eq:P2}) we have assumed for the NSI coupling the general complex form
\begin{eqnarray}
\varepsilon_{\alpha \beta} = |\varepsilon_{\alpha \beta}|  e^{i\phi_{\alpha \beta}}\,.
\end{eqnarray}
The expression of $P_2$ is different for $\varepsilon_{e\mu}$ and  $\varepsilon_{e\tau}$ and,
in Eq. (\ref{eq:P2}), one has to make the replacements
\begin{eqnarray}
 \label{eq:P2_NSI_1}
 a = s^2_{23}, \quad b = c^2_{23} \quad &{\mathrm {if}}& \quad \varepsilon = |\varepsilon_{e\mu}|e^{i{\phi_{e\mu}}}\,,\\
  \label{eq:P2_NSI_2}
 a =  s_{23}c_{23}, \quad b = -s_{23} c_{23} \quad &{\mathrm {if}}& \quad \varepsilon = |\varepsilon_{e\tau}|e^{i{\phi_{e\tau}}}\,.
\end{eqnarray}
In the expressions given in Eqs.~(\ref{eq:P0})-(\ref{eq:P2}) for $P_0$, $P_1$ and $P_2$, 
the sign of $\Delta$, $\alpha$ and $v$ is positive (negative) for NO (IO).  
We recall that the expressions of the probability provided above hold for
neutrinos and that the  corresponding formulae for antineutrinos can be derived
by flipping in Eqs.~(\ref{eq:P0})-(\ref{eq:P2}) the sign of all the CP-phases and of the matter parameter $v$. 
Finally, we observe that the third term $P_{\rm {2}}$ encodes the dependency
on the (complex) NSI coupling and it is different from zero only in matter (i.e. if $v \ne 0$). 
It is generated by the interference of the matter potential
 $\varepsilon_{e\mu}V_{CC}$  (or $\varepsilon_{e\tau}V_{CC}$)
 with the atmospheric wavenumber $k_{31}$ (see the discussion in~\cite{Friedland:2012tq})%
\footnote{Interestingly, an analogous splitting $P_{\mu e}  \simeq  P_{\rm{0}} + P_{\rm {1}}+   P_{\rm {2}}$
of the transition probability is valid in the presence of 
oscillations driven by a sterile neutrino~\cite{Klop:2014ima}. In that case, however, the term $P_2$ 
emerges due to the interference between the amplitude driven by the
atmospheric mass difference and that by the mass difference
corresponding to the sterile neutrino, instead of the interference
with the term originated from the matter potential.}.

{\bf {\em  Data used in the analysis.}}  We extracted the datasets of NO$\nu$A and T2K from the latest data released 
  in~\cite{NOVA_talk_nu2020} and~\cite{T2K_talk_nu2020}.
  We fully incorporate  both the disappearance and appearance channels in both experiments. 
 In our analysis we use the software GLoBES~\cite{Huber:2004ka,Huber:2007ji} 
and its additional public tool~\cite{Kopp:NSI}, which can implement NSI.
In our analysis we have marginalized over $\theta_{13}$  with 3.4\% 1 sigma prior with central value 
$\sin^2\theta_{13}= 0.0219$ as determined by Daya Bay~\cite{Adey:2018zwh}.
 We have fixed the solar parameters $\Delta m^2_{21}$ and $\theta_{12}$
at their best fit values estimated in the recent global analysis~\cite{Capozzi:2017ipn}.

{\bf {\em Numerical Results.}} Figure~\ref{fig:regions_1} reports the results of the analysis 
of the combination of T2K and NO$\nu$A for NO (left panels) and IO (right panels). 
The upper (lower) panels refer to $\varepsilon_{e\mu}(\varepsilon_{e\tau}$)
taken one at a time. Each panel displays the allowed regions in the plane spanned by
the relevant NSI coupling and the standard CP-phase $\delta_{\mathrm {CP}}$. 
The non-standard CP-phases, the mixing angles $\theta_{23}$ and $\theta_{13}$, and
the squared-mass $\Delta m^2_{31}$ are marginalized away. We display the allowed regions at the
68\% and 90\% confidence level for 2 d.o.f., and denote with a star the best fit point.
From the left upper panel we can appreciate that in NO there is a $\sim$ 2.1$\sigma$ ($\Delta \chi^2 = 4.50$) preference 
for a non-zero value of the coupling $|\varepsilon_{e\mu}|$, with best fit $|\varepsilon_{e\mu}| =0.15$.
In the right upper panel we see that in IO the preference for NSI is negligible.
The lower panels depict the situation for the coupling $|\varepsilon_{e\tau}|$. 
In NO there is a preference at the 1.9$\sigma$ ($\Delta \chi^2 = 3.75$) with best fit $|\varepsilon_{e\tau}| =0.27$, 
while in IO the preference is only at the 1.0$\sigma$ with best fit $|\varepsilon_{e\tau}| =0.15$.
It is interesting to note how in all four cases the preferred value for the CP-phase $\delta_{\mathrm {CP}}$ is close to
$3\pi/2$. We will come back later on this important point. 

Figure~\ref{fig:regions_2} shows the results of the analysis of the combination of T2K and NO$\nu$A
similar to Fig.~\ref{fig:regions_1}. In this case, however, each panel displays the allowed regions in the 
plane spanned by the relevant NSI coupling ($|\varepsilon_{e\mu}|$ or $|\varepsilon_{e\tau}|$) and the 
corresponding CP-phase ($\phi_{e\mu}$ or $\phi_{e\tau}$). The standard CP-phase $\delta_{\mathrm {CP}}$, 
the mixing angles $\theta_{23}$ and $\theta_{13}$, and
the squared-mass $\Delta m^2_{31}$ are marginalized away. 
It is intriguing to note how in the NO case the preferred value for both the new CP-phases 
$\phi_{e\mu}$ and $\phi_{e\tau}$ is close to $3\pi/2$, so indicating purely imaginary NSI,
i.e. maximal CP-violation also in the NSI sector. In Table~\ref{table:chi2} we report the best fit values of the NSI couplings
together with the CP-phases and the value of $\Delta\chi^2=\chi^2_{\rm SM}-\chi^2_{\rm SM + NSI}$ 
for a fixed choice of the NMO.

\begin{figure}[t!]
\vspace*{-0.0cm}
\hspace*{-0.2cm}
\includegraphics[height=4.2cm,width=4.2cm]{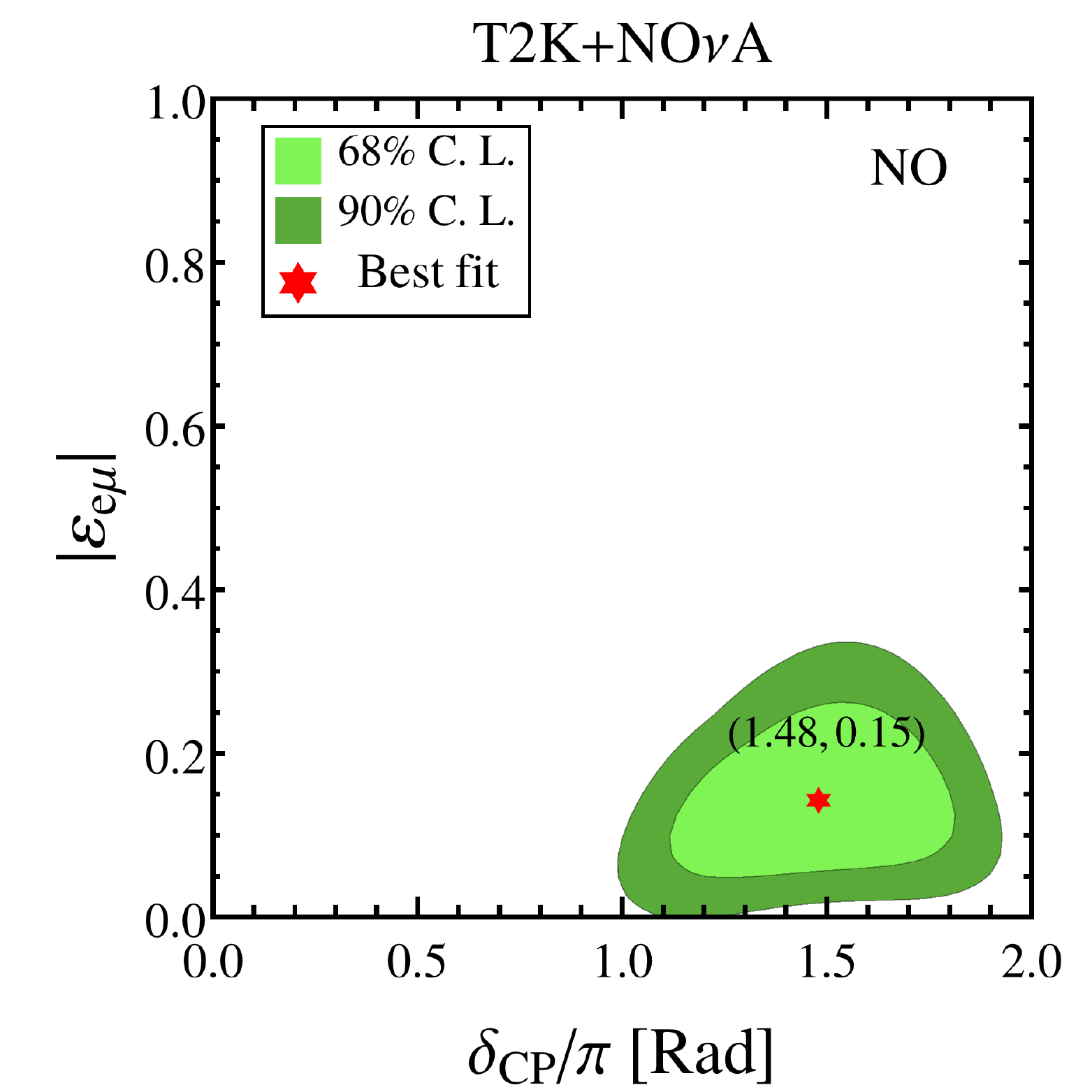}
\includegraphics[height=4.2cm,width=4.2cm]{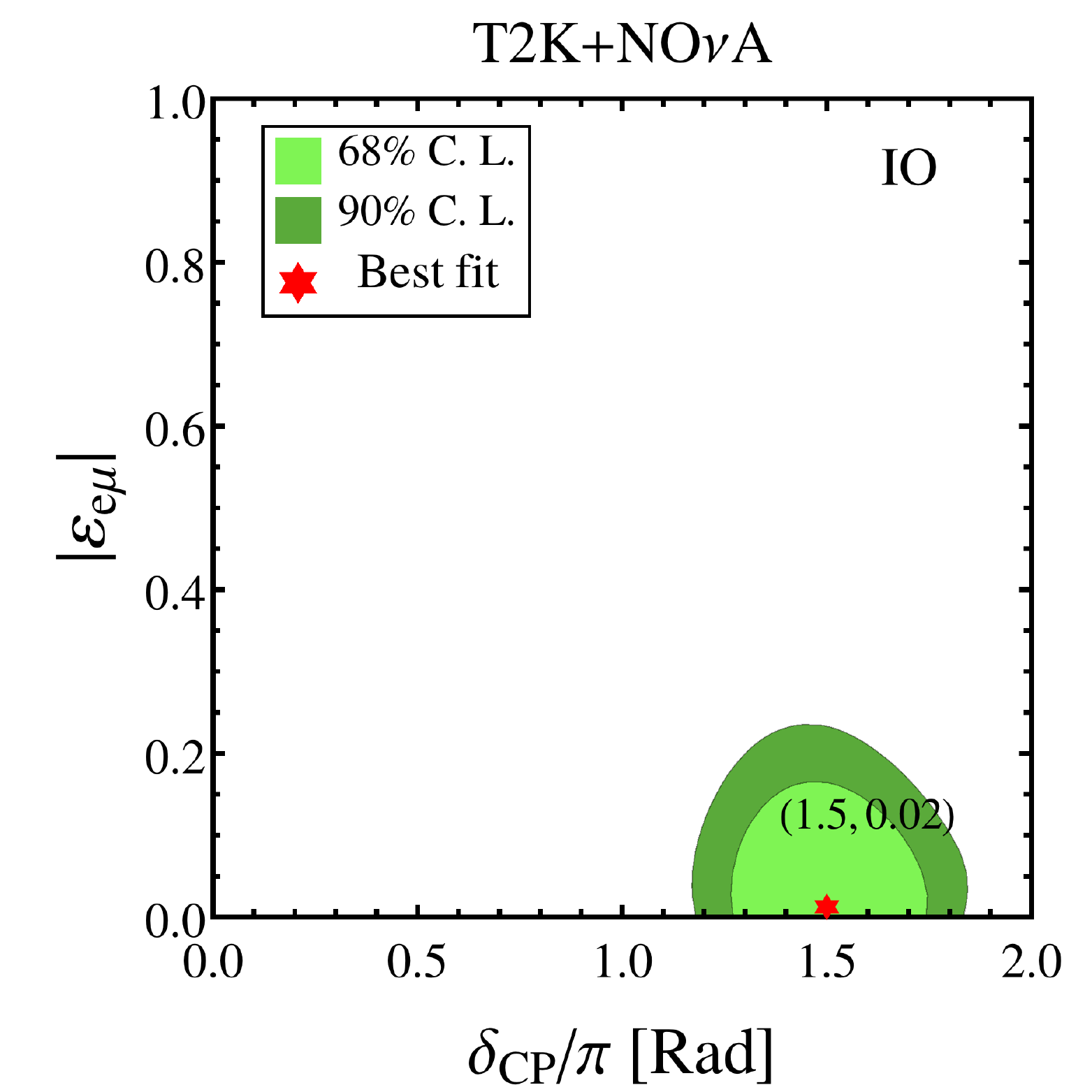}
\includegraphics[height=4.2cm,width=4.2cm]{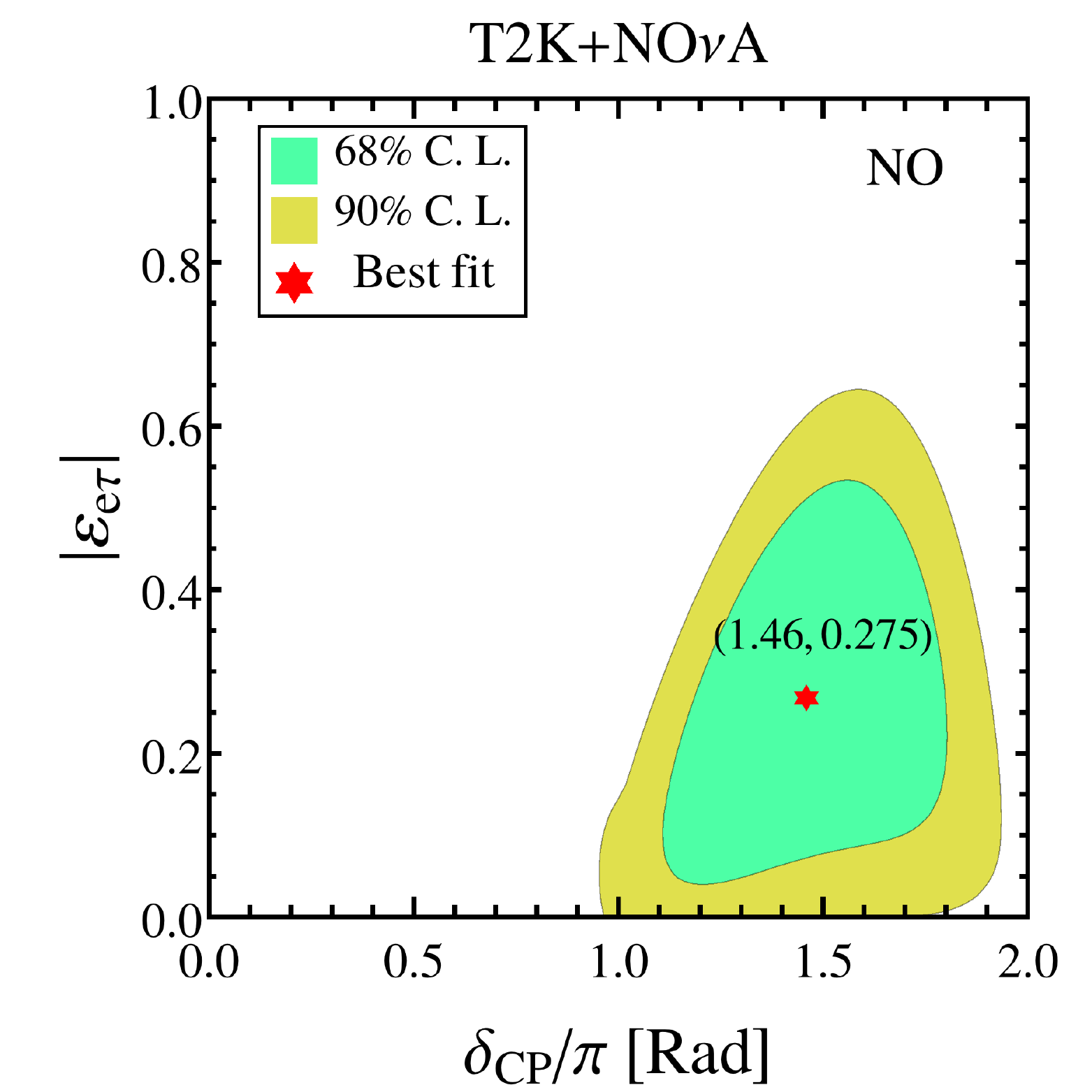}
\includegraphics[height=4.2cm,width=4.2cm]{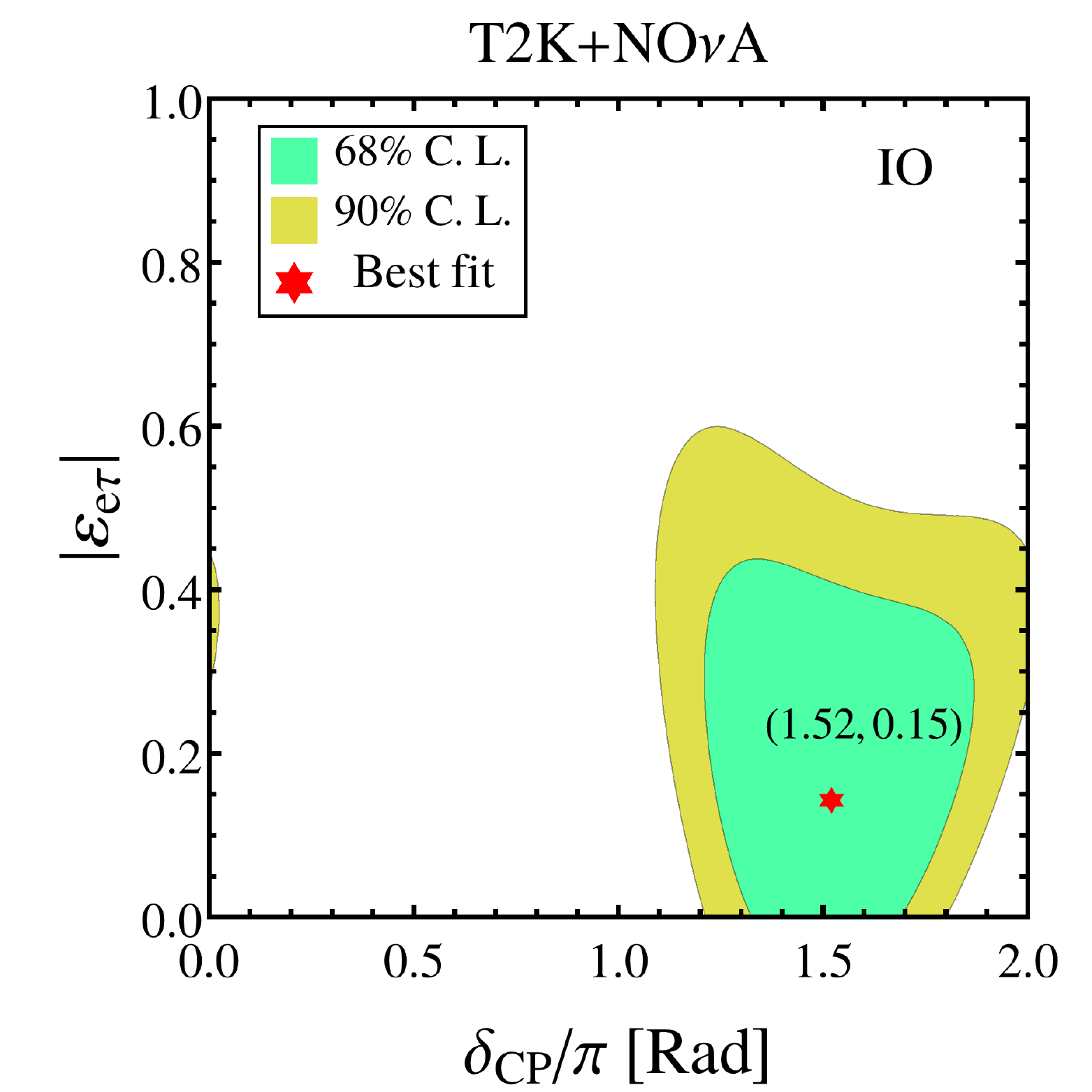}
\caption{Allowed regions determined by the combination of T2K and NO$\nu$A for NO (left panels)
and IO (right panels). The upper (lower) panels refer to $\varepsilon_{e\mu}(\varepsilon_{e\tau}$)
taken one at a time. In the upper (lower) panels the NSI CP-phase $\phi_{e\mu}$ ($\phi_{e\tau}$) 
has been marginalized. In all panels the atmospheric parameters $\Delta m^2_{31}$ and $\theta_{23}$ 
have been marginalized. The contours are drawn at the 68\% and 90\% confidence level for 2 d.o.f..}
\label{fig:regions_1}
\end{figure} 

\begin{table}[b!]
\centering
\caption{Best fit values and $\Delta\chi^2=\chi^2_{\rm SM}-\chi^2_{\rm SM+NSI}$ for the two choices of the NMO.}
\vspace{0.5cm}
\begin{tabular}{c|c|c|c|c|c}
NMO&NSI&$|\varepsilon_{\alpha\beta}|$&$\phi_{\alpha\beta}/\pi$&$\delta_{\mathrm {CP}}/\pi$&$\Delta\chi^2$\\\hline
\multirow{2}{*}{NO}&$\varepsilon_{e\mu}$&0.15&1.38&1.48&4.50\\
&$\varepsilon_{e\tau}$&0.27&1.62&1.46&3.75\\\hline
\multirow{2}{*}{IO}&$\varepsilon_{e\mu}$&0.02&0.96&1.50&0.07\\
&$\varepsilon_{e\tau}$&0.15&1.58&1.52&1.01\\\hline
\end{tabular}
\label{table:chi2}
\end{table}

\begin{figure}[t!]
\vspace*{-0.0cm}
\hspace*{-0.2cm}
\includegraphics[height=4.2cm,width=4.2cm]{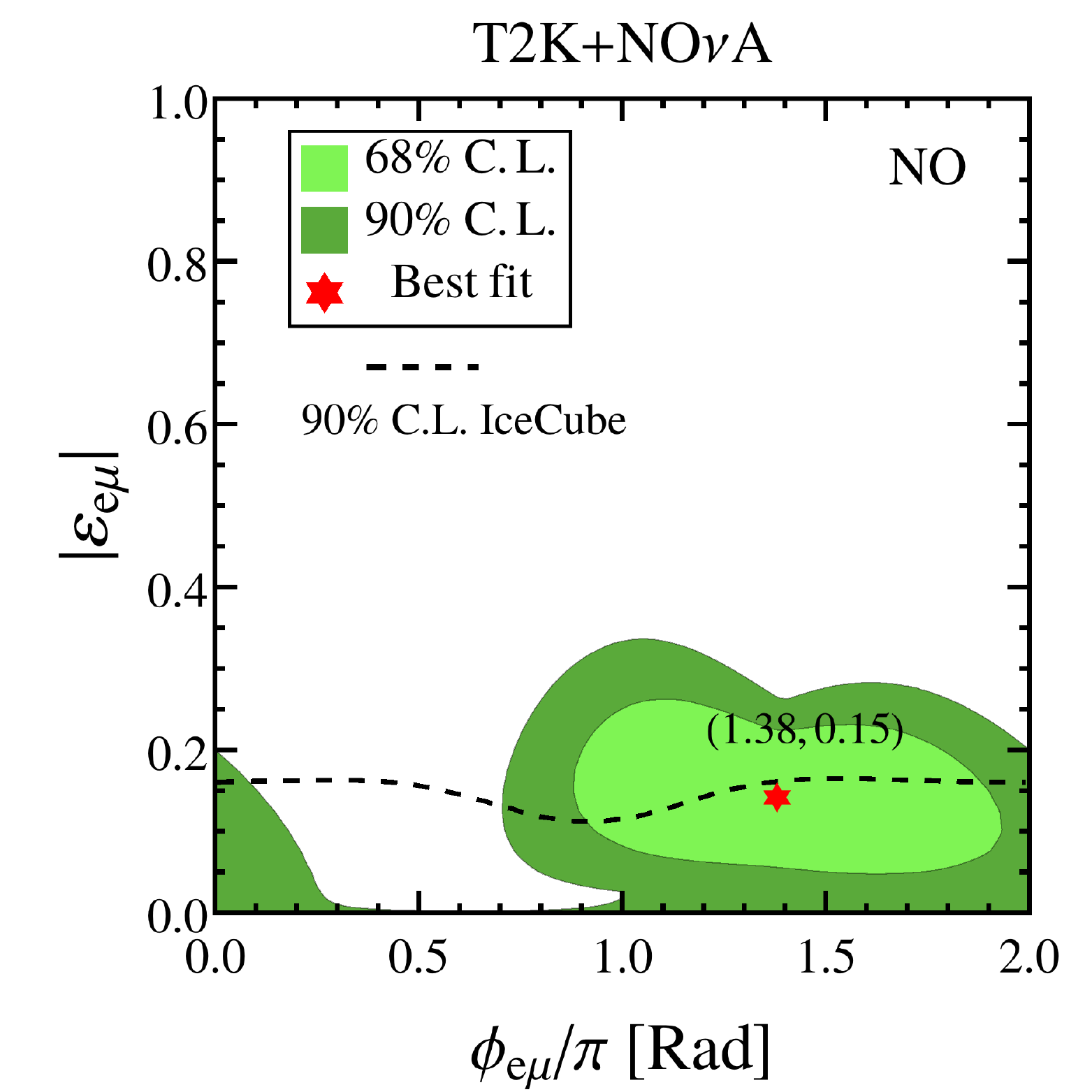}
\includegraphics[height=4.2cm,width=4.2cm]{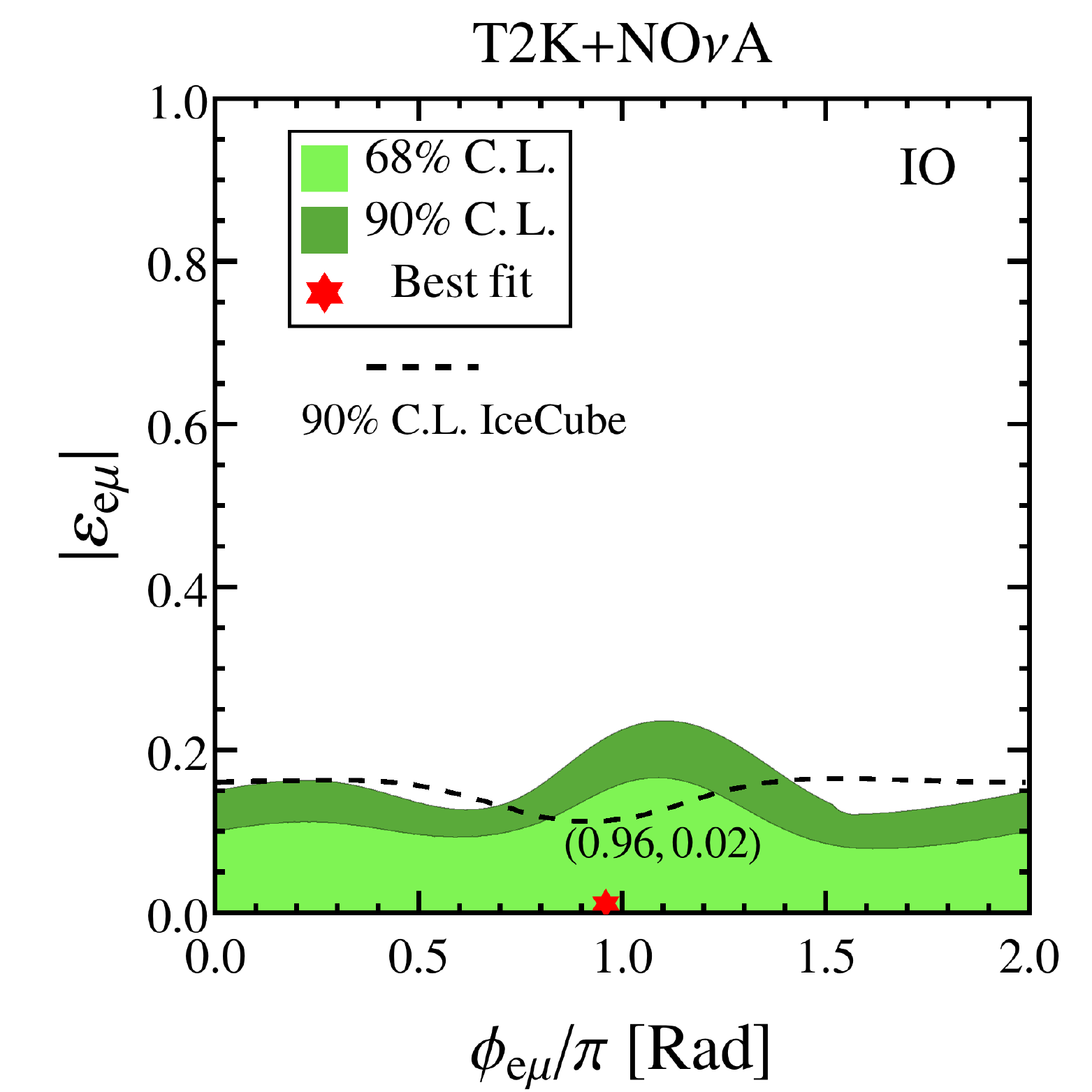}
\includegraphics[height=4.2cm,width=4.2cm]{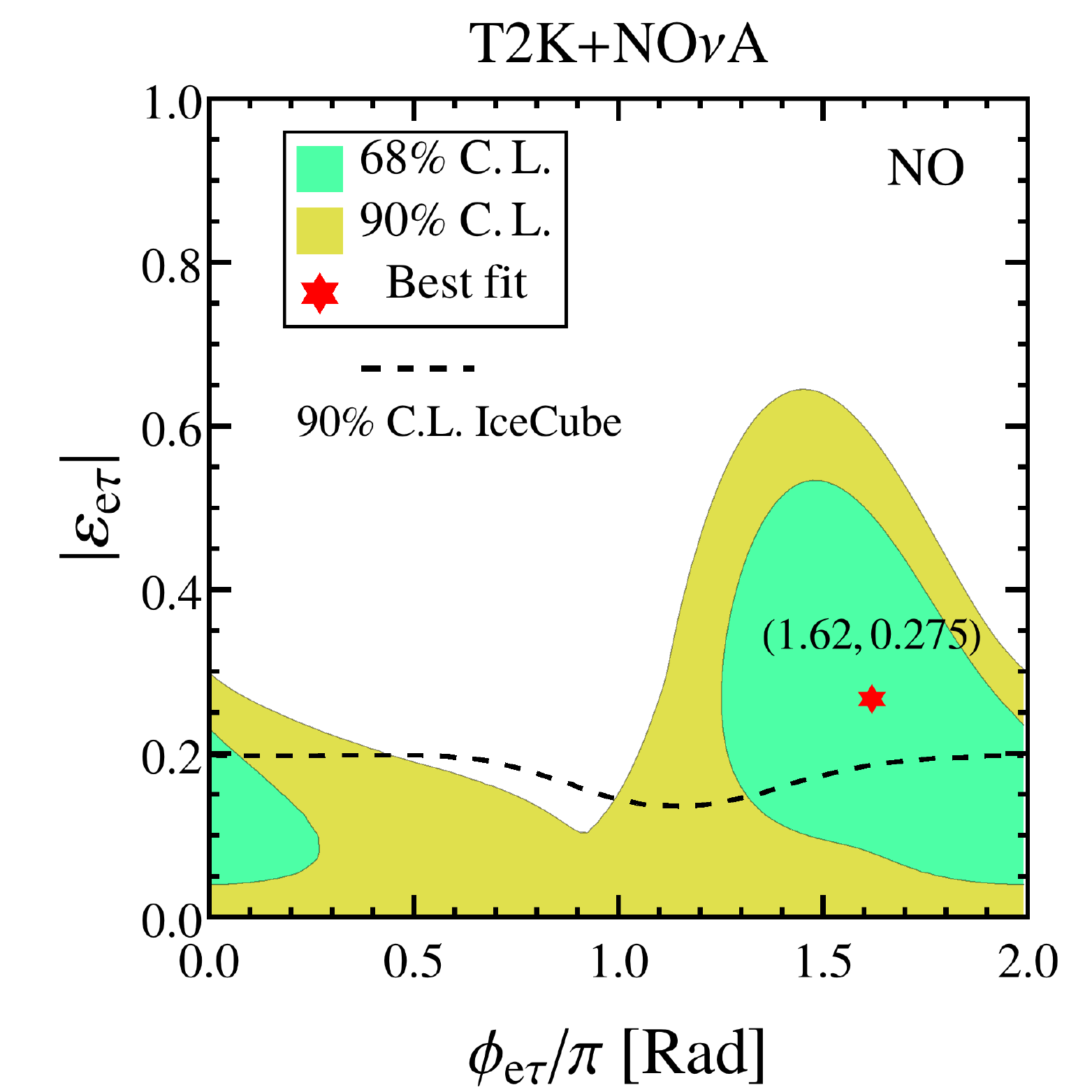}
\includegraphics[height=4.2cm,width=4.2cm]{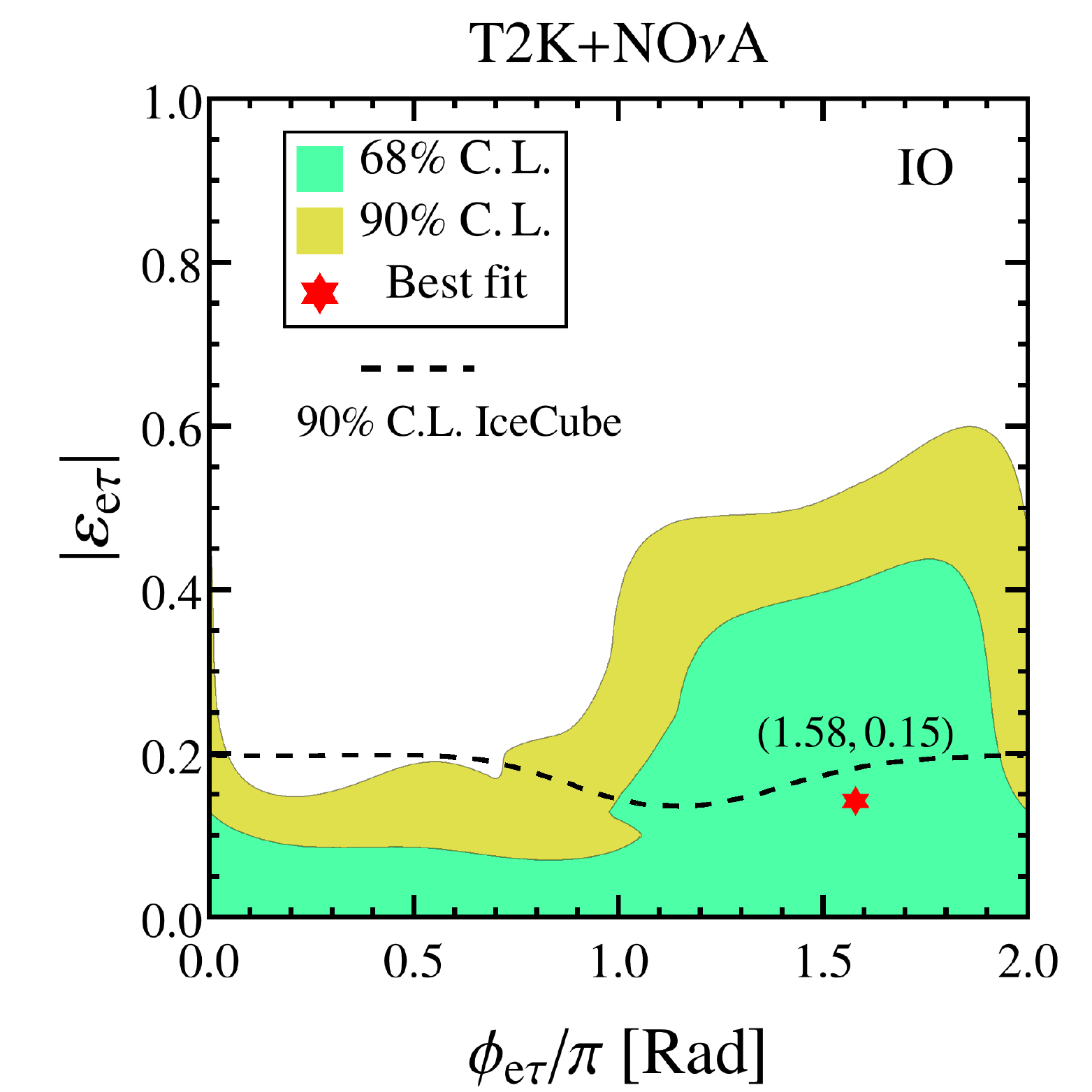}
\caption{Allowed regions determined by the combination of T2K and NO$\nu$A for NO (left panels)
and IO (right panels). The upper (lower) panels refer to $\varepsilon_{e\mu}(\varepsilon_{e\tau}$)
taken one at a time. In all panels the standard CP-phase $\delta_{\rm CP}$ 
has been marginalized in addition to the atmospheric parameters $\Delta m^2_{31}$ and $\theta_{23}$.
The contours are drawn at the 68\% and 90\% confidence level for 2 d.o.f..
The dashed curves represent the upper bounds (90\% C.L., 2 d.o.f.) derived from the preliminary 
analysis of the IceCube data~\cite{IceCube_talk_PPNT2020}.}
\label{fig:regions_2}
\end{figure} 

In Fig.~\ref{fig:regions_2}, we superimpose
the upper bounds coming from the preliminary analysis of IceCube data~\cite{IceCube_talk_PPNT2020},
which are the most stringent ones in the literature on the relevant couplings.
These bounds are not incompatible with the indication we find.
Rather, they select the lower values of the couplings favored by T2K and NO$\nu$A. Interestingly, IceCube finds 
$|\varepsilon_{e\mu}| = 0.07$ as best fit point with a preference of 1 sigma level with respect to the SM case (see slides 20 and 33
in~\cite{IceCube_talk_PPNT2020}). Also, the best fit we find for the CP-phase $\phi_{e\mu} \sim 3\pi/2$ is compatible with that found by IceCube. Although we cannot quantitatively combine our results with those of IceCube, we can estimate
that $|\varepsilon_{e\mu}| \sim 0.1$  is expected to come as the best fit 
from such a combination with a significance around the 2 sigma confidence level.

\begin{figure*}[t!]
\vspace*{-0.0cm}
\hspace*{-0.1cm}
\includegraphics[height=5.87cm,width=5.87cm]{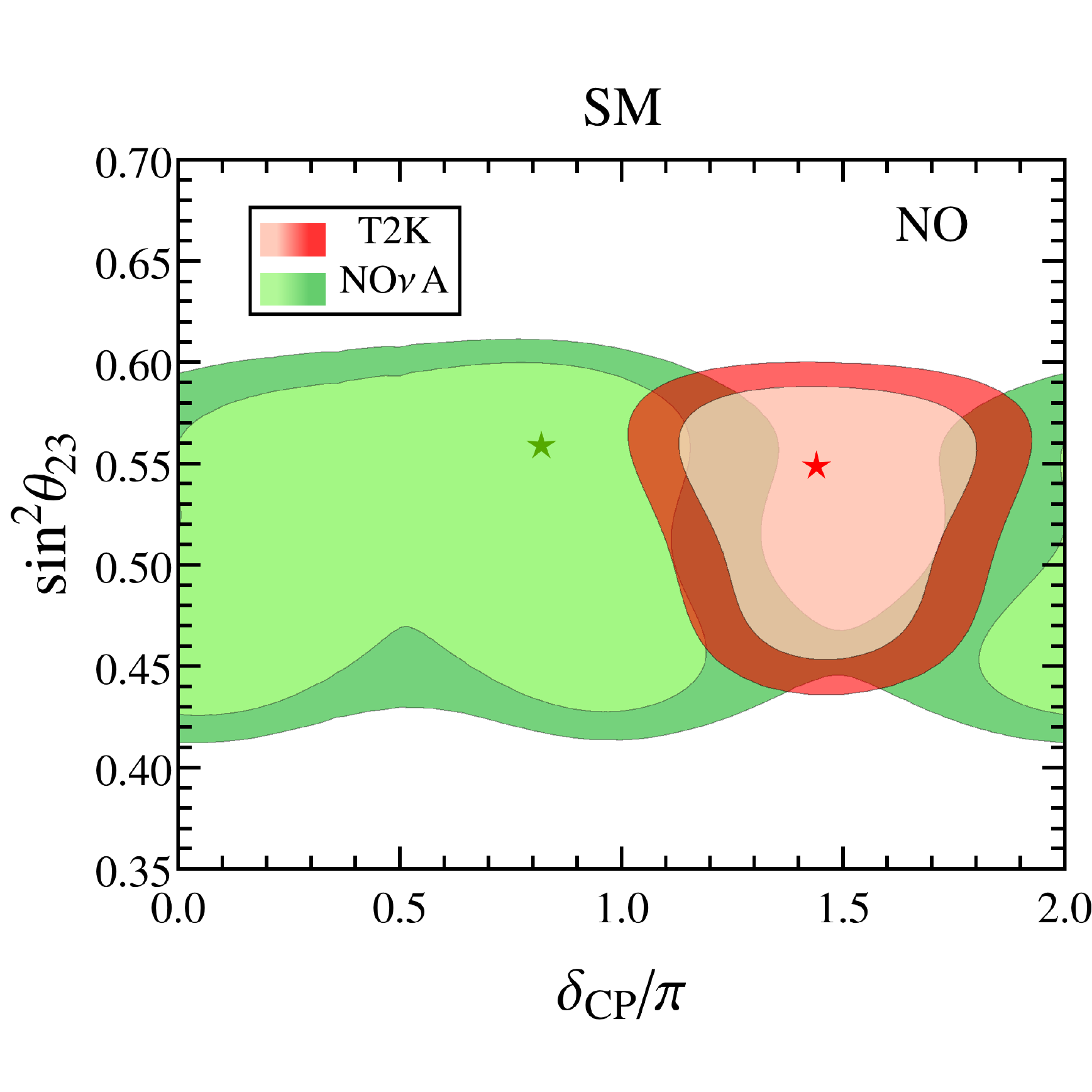}
\includegraphics[height=5.87cm,width=5.87cm]{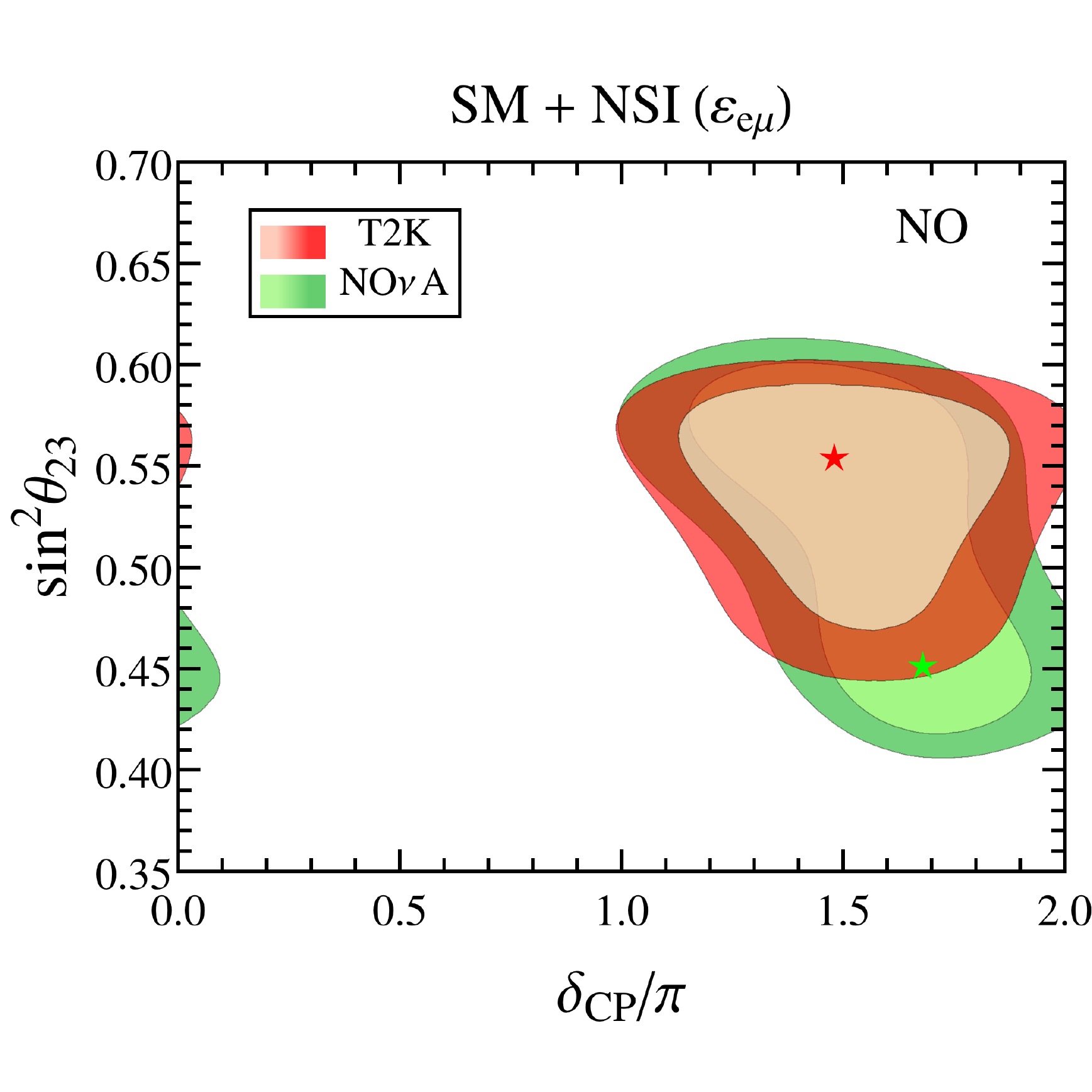}
\includegraphics[height=5.87cm,width=5.87cm]{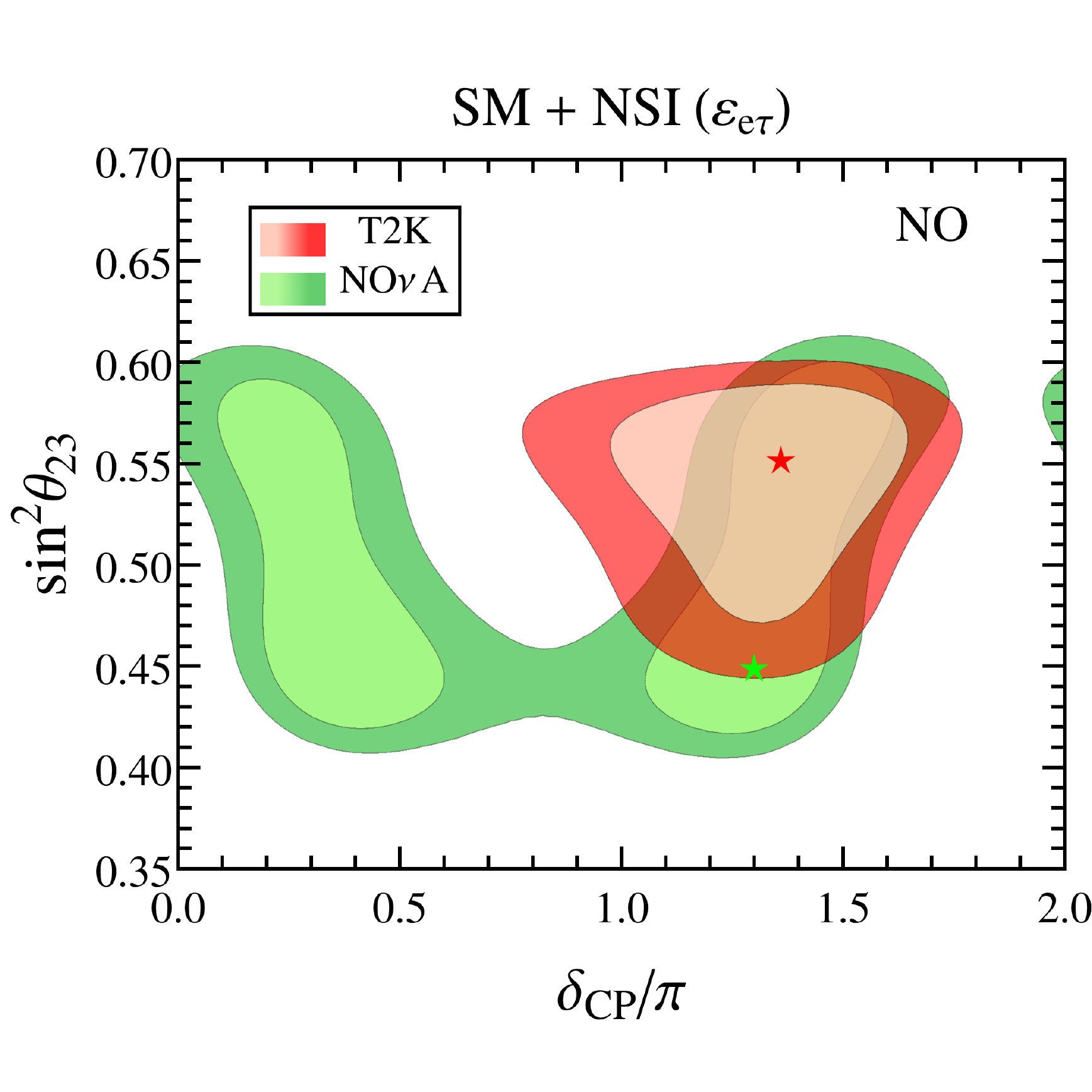}
\vspace*{-0.3cm}
\caption{Allowed regions determined separately by T2K and NO$\nu$A for NO in the SM case
(left panel) and with NSI in the $e-\mu$ sector (middle panel) and in the $e-\tau$ sector
 (right panel). In the middle panel we have taken 
the NSI parameters at their best fit values of T2K + NO$\nu$A ($|\varepsilon_{e\mu}| = 0.15, \phi_{e\mu} = 1.38\pi$).
Similarly, in the right panel we have taken $|\varepsilon_{e\tau}| = 0.275, \phi_{e\tau} = 1.62\pi$.
The contours are drawn at the 68$\%$ and 90$\%$ C.L. for 2 d.o.f. The comparison of the  middle and right panels
with the left one clearly evidences the reduction of the tension between the two experiments in the presence of NSI
of both types.}
\label{fig_tension}
\end{figure*} 

In order to understand how the preference for a non-zero NSI coupling arises, it is useful to look
to what happens separately to NO$\nu$A and T2K. For this purpose, in 
Fig.~\ref{fig_tension} we display the  allowed regions in the 
plane spanned by the standard CP-phase $\delta_{\mathrm {CP}}$ and the atmospheric 
mixing angle $\theta_{23}$ in the NO case. The left panel refers to the SM case,
while the middle and right panels concern the SM+NSI scenario with NSI in the $e-\mu$ and $e-\tau$ sectors
respectively.
In the middle and right panels we have taken the NSI parameters at their best fit values of the combined
analysis of NO$\nu$A and T2K. More specifically, $|\varepsilon_{e\mu}| = 0.15, \phi_{e\mu} = 1.38\pi$ (middle panel)
and $|\varepsilon_{e\tau}| = 0.275, \phi_{e\tau} = 1.62\pi$ (right panel). The contours are drawn at the 68$\%$ and 90$\%$ C.L. for 2 d.o.f. 
In the SM case a clear mismatch in the determination of the CP-phase $\delta_{\mathrm {CP}}$ among the two experiments is
evident. While NO$\nu$A prefers values close to $\delta_{\mathrm {CP}} \sim 0.8 \pi$, T2K identifies a value of $\delta_{\mathrm {CP}} \sim 1.4 \pi$.
Such two estimates, which have a difference of phase of about $\pi/2$, are in disagreement at more than  90$\%$ C.L. for 2 d.o.f..
The reduction of the tension between the two experiments obtained in the presence of NSI is evident both in the middle and right panels where the best fit values of $\delta_{\mathrm {CP}}$ are very close to the common value  $\delta_{\mathrm {CP}} \sim 3\pi/2$. We see that the value of $\delta_{\mathrm {CP}}$ preferred by T2K is basically unchanged in the presence of NSI as this experiment has a reduced sensitivity to matter effects. As a consequence the value of  $\delta_{\mathrm {CP}}^{\mathrm {T2K}}  \sim 3\pi/2$ identified by T2K can be considered a faithful estimate of its true value both in SM and in SM+NSI scenarios. In contrast, NO$\nu$A due to the enhanced sensitivity to matter effects, if NSI are not taken into account (left panel), identifies a fake value of  
$\delta_{\mathrm {CP}}^{\mathrm {NOvA}}  \sim 0.8\pi$. In NO$\nu$A, the preference for the true value of $\delta_{\mathrm {CP}}  \sim 3\pi/2$ is restored once the NSI are taken into account (middle and right panels). Therefore, it seems that NSI offer a very simple and elegant way to solve the discrepancy among the two experiments. We also note that the allowed regions
for NO$\nu$A are qualitatively different in the $e-\mu$ and $e-\tau$ NSI cases. In fact, in the first case there is a single
allowed region while in the second case there are two degenerate lobes. This different behavior can be traced
to the fact that the transition probabilities are different in the two cases. More specifically, the  sign in front of the coefficient
$b$ of $P_2$ in Eq.~(\ref{eq:P2}) [see Eqs.~(\ref{eq:P2_NSI_1}) and (\ref{eq:P2_NSI_2})] is opposite in the two scenarios.

For completeness, in the  Supplemental Material (which includes references~\cite{Kelly:2020fkv,Esteban:2020cvm,Denton:2020uda}),
we provide three additional figures. First, we present a so-called bievents plot (Fig.~S1) meant to elucidate the tension between T2K
and NO$\nu$A and its resolutions with NSI. Second, we provide a figure (Fig.~S2) analogous to Fig.~\ref{fig_tension} but referring
to the IO case. This plot clarifies why (as shown in Figs.~\ref{fig:regions_1} and \ref{fig:regions_2} and also in Table~\ref{table:chi2}), 
in the IO case there is basically no preference for non-zero NSI.  Finally, in Fig.~S3, we show the one-dimensional projections on the standard 
oscillation parameters $\delta_{\mathrm {CP}}$, $\theta_{23}$ and $|\Delta m^2_{31}|$
from the combination of NO$\nu$A and T2K, with and without NSI. Note that Fig. S1 is not present in the Supplemental Material
published in PRL.

{\bf {\em Conclusions.}} In this paper we have investigated the impact of NSI on
the tension recently emerged in the latest T2K and NO$\nu$A data. Our main result is that 
such a tension can be resolved by non-standard interactions (NSI) of the flavor changing type involving the
$e-\mu$ and $e-\tau$ flavors. We underline that, apart from the LBL accelerator data, 
it would be very important to complement our study considering the 
atmospheric neutrino data. To this regard, we mention the recent IceCube DeepCore 
analysis~\cite{IceCube_talk_PPNT2020}, which starts
to probe values of the NSI couplings below $\sim0.2$, close but not incompatible 
to those relevant to the present analysis. We also hope
that SuperKamiokande may provide an updated analysis of the atmospheric data in the presence
of NSI, which is currently unfeasible from outside the collaboration. 

Our results point towards relatively large effective NSI couplings of the order of ten per cent. Taking into
 account Eq.~(\ref{epsilon_eff}), these may correspond to couplings of a few per 
cent at the level of the fundamental constituents (u and d quarks and electrons). 
A major challenge in generating such observable NSI in any UV-complete model is that there are stringent bounds
arising from charged-lepton flavor violation. In fact, when the new physics responsible 
for the generation of the neutrino NSI is due to mediators heavier than the electroweak 
symmetry-breaking scale, one expects also that the charged leptons are involved as components
of the doublet of SU(2). One possible way to circumvent this problem is to increase the complexity
of the model. At the tree-level, one can consider NSI generated in models endowed with 
dimension-8 operators, which typically require the introduction of two mediators~\cite{Gavela:2008ra}.
Another possibility, remaining in the framework of heavy-mediators induced NSI, is to consider NSI which
arise in radiative neutrino mass models (see for example the recent studies~\cite{Babu:2019mfe,Forero:2016ghr,Dey:2018yht}).
A third possibility is to abandon altogether the heavy-mediator paradigm and consider NSI induced by light mediators (see for example~\cite{Farzan:2015doa,Farzan:2015hkd}.).

In this manuscript we have focused on the current data provided by NO$\nu$A and T2K.
Needless to say, it would be interesting to consider the sensitivity to NSI of the future LBL experiments.
In particular, we foresee that a careful comparison of T2HK and DUNE should be very informative.
On the one hand T2HK, with its short 295 km baseline should be able to determine the standard parameters
almost independently of NSI. On the other hand, DUNE with its 1300 km baseline should manifest 
striking effects induced by NSI and allow their identification. Of course, the determination
of the NMO is expected to become more challenging in the presence of new physics.
To this respect we underline the importance of experiments like JUNO which are insensitive to 
(both standard and non-standard) matter effects and will allow us to identify 
the NMO (and  other standard oscillation parameters) independently of hypothetical NSI.
Finally, we note that independent measurements of the NSI couplings relevant for NO$\nu$A and T2K may 
also come in the future from experiments that probe the coherent elastic neutrino nucleus scattering. 


{\bf {\em Note.}} In the final stage of preparation of our manuscript the paper~\cite{Denton:2020uda} 
appeared discussing a similar scenario. 

\begin{acknowledgments}

\noindent  A.P. acknowledges partial support by the research grant number 2017W4HA7S ``NAT-NET: Neutrino and Astroparticle Theory Network'' under the program PRIN 2017 funded by the Italian Ministero dell'Istruzione, dell'Universit\`a e della Ricerca (MIUR) and by the research project {\em TAsP} funded by the Instituto Nazionale di Fisica Nucleare (INFN). 

\end{acknowledgments}

\bibliographystyle{h-physrev41}
\bibliography{NSI-References_v2}

\clearpage
\newpage
\maketitle
\onecolumngrid
\begin{center}
\textbf{\large Supplemental Material} \\ 
\vspace{0.05in}

\end{center}

\setcounter{equation}{0}
\setcounter{figure}{0}
\setcounter{table}{0}
\setcounter{section}{1}
\renewcommand{\theequation}{S\arabic{equation}}
\renewcommand{\thefigure}{S\arabic{figure}}
\renewcommand{\thetable}{S\arabic{table}}
\newcommand\ptwiddle[1]{\mathord{\mathop{#1}\limits^{\scriptscriptstyle(\sim)}}}

\begin{figure}[b!]
\vspace*{-0.0cm}
\hspace*{-0.2cm}
\includegraphics[height=6.2cm,width=6.2cm]{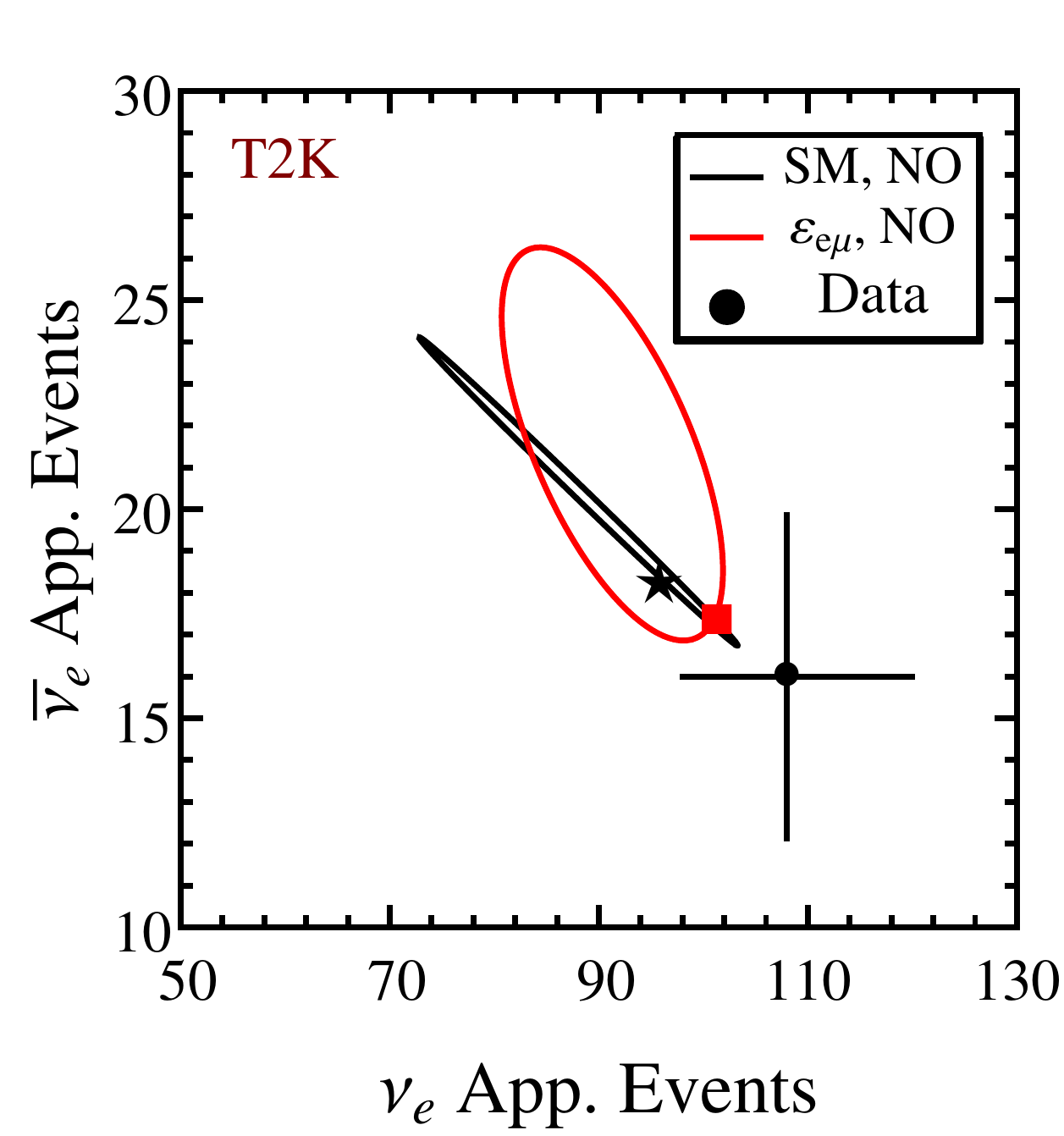}
\includegraphics[height=6.2cm,width=6.2cm]{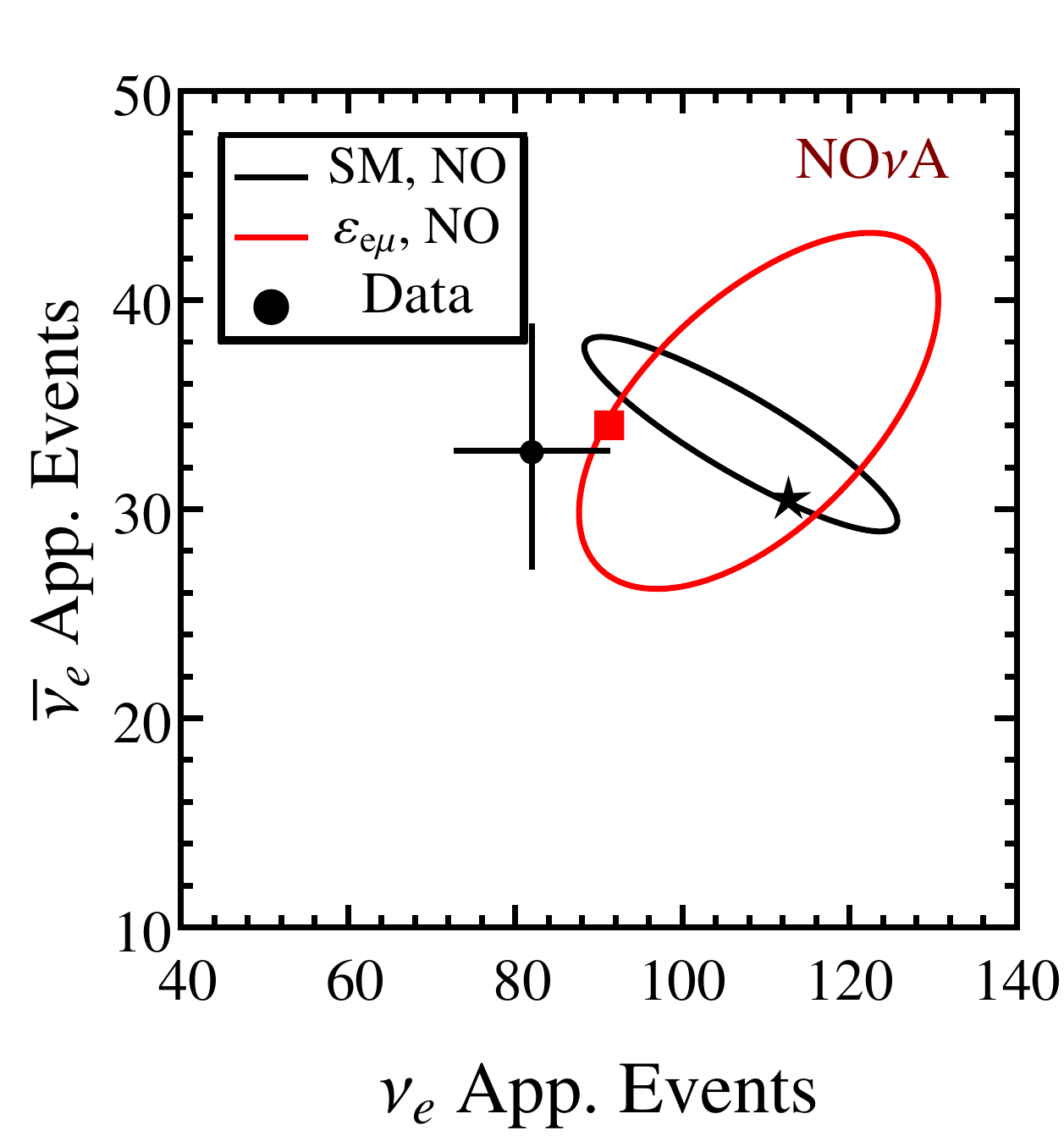}
\includegraphics[height=6.2cm,width=6.2cm]{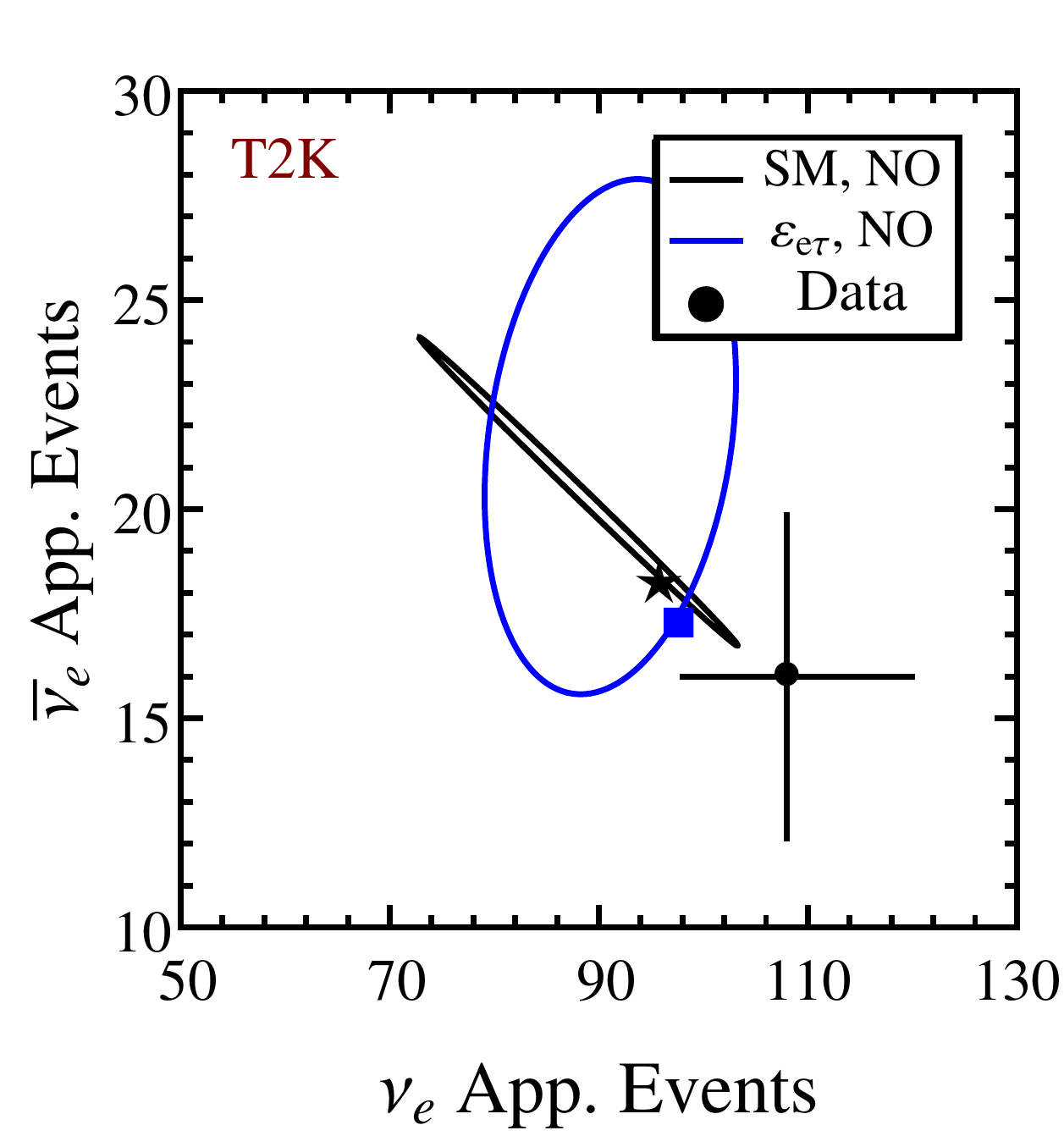}
\includegraphics[height=6.2cm,width=6.2cm]{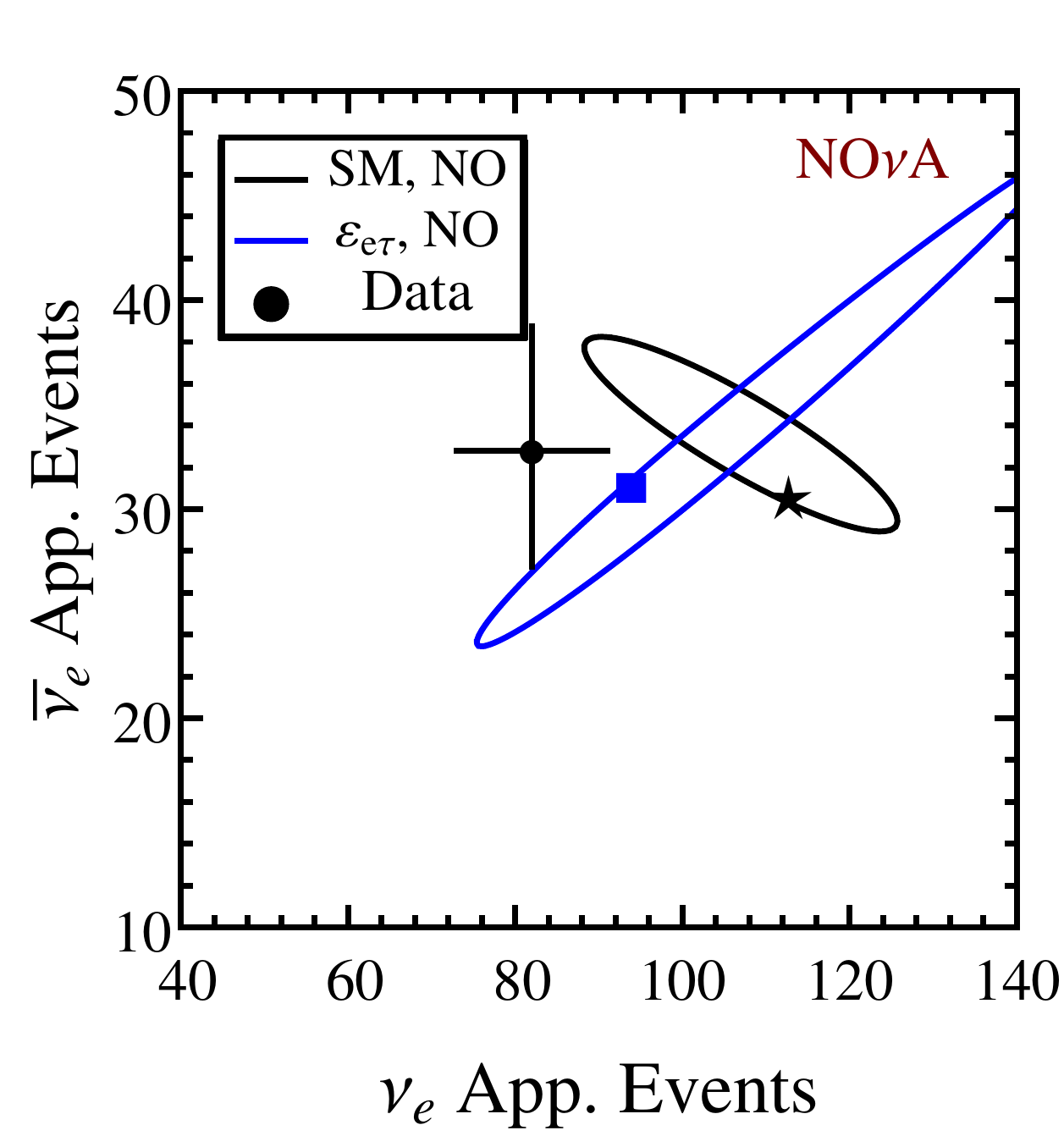}
\caption{Bievents plot for the T2K (left panels) and NO$\nu$A setup (right panels). 
The upper (lower) panels illustrate the case of $\varepsilon_{e\mu} (\varepsilon_{e\tau}$).
In all the ellipses the varying parameter is the standard CP phase $\delta_{\rm CP}$ in the
range $[0, 2\pi]$. The black ellipses represent the SM case with best fit represented by stars.
The colored ellipses represent the SM+NSI case with best fit indicated by squares. 
The ellipses and the best fit points located on them are determined by fitting the {\em combination} 
of the two experiments T2K and NO$\nu$A. The points with the error bars represent the experimental 
data with their statistical uncertainties.}
\label{fig:bievents-plot}
\end{figure}

\noindent In this Supplemental Material, we provide and discuss three additional figures. In Fig.~\ref{fig:bievents-plot},
we present a so-called bievents plot, representing on the $x$-axis ($y$-axis) the number of electron neutrino (antineutrino)
 events measured in the appearance channel. This figure helps to understand the origin of the discrepancy of T2K and NO$\nu$A
found in the SM scenario in the NO case, and its resolution in the presence of NSI. This kind of plot
is very useful because in NO$\nu$A and T2K, essentially,  all the information 
provided by the appearance channel measurements is contained in the number of events collected. In fact, due to the limited statistics, 
the information extracted from the shape of the energy spectrum is currently very limited. Therefore, although we have included
in our numerical analysis the full binned energy spectrum, considerations based on the number of events can be 
effectively used to understand the numerical results.
The ellipses shown in the plots are obtained from the {\em combination} of T2K and NOvA. In the SM case, 
the ellipses correspond to the best fit values of the oscillation parameters $\theta_{23}$, $\theta_{13}$ and $\Delta m^2_{31}$. 
In the SM+NSI case, the ellipses correspond to the best fit values of $\theta_{23}$, $\theta_{13}$, $\Delta m^2_{31}$, $|\varepsilon_{\alpha\beta}|$ 
and $\phi_{\alpha\beta}$. Both in the SM and SM+NSI cases, the varying parameter on the ellipses is the standard CP 
phase $\delta_{\rm CP}$ in the range $[0, 2\pi]$. The black ellipses represent the SM case 
(the stars indicate the best fit value $\delta_{\rm CP}^{\rm SM} = 1.16\pi$). This value of the CP phase is obtained
as a compromise between T2K (which strongly pushes towards  $\delta_{\rm CP} = 1.5\pi$) and NO$\nu$A (which 
tends to prefer values close to $0.8\pi$). It is clear that the best fit point of the SM is not in good agreement 
 with the experimental data of both experiments, implying the tension observed in the analysis.
The colored ellipses represent the SM+NSI case
 (the squares indicate the best fit value of $\delta_{\rm CP}^{\rm NSI} \simeq 1.5 \pi$, in both the $e-\mu$ and $e-\tau$ cases). 
 The upper (lower) panels refer to the $e-\mu$ ($e-\tau$) case. In the presence of NSI, the NO$\nu$A ellipses are completely
 different from those obtained for T2K, because the impact of the matter effects is different (they are much larger in NO$\nu$A).
 From the figure, it clearly emerges how in the presence of the NSI, the best fit point of the model is very close
 to the experimental data of both experiments, thus resolving the discrepancy observed in the SM case.
 


\begin{figure*}[h!]
\vspace*{-0.0cm}
\hspace*{-0.1cm}
\includegraphics[height=5.87cm,width=5.87cm]{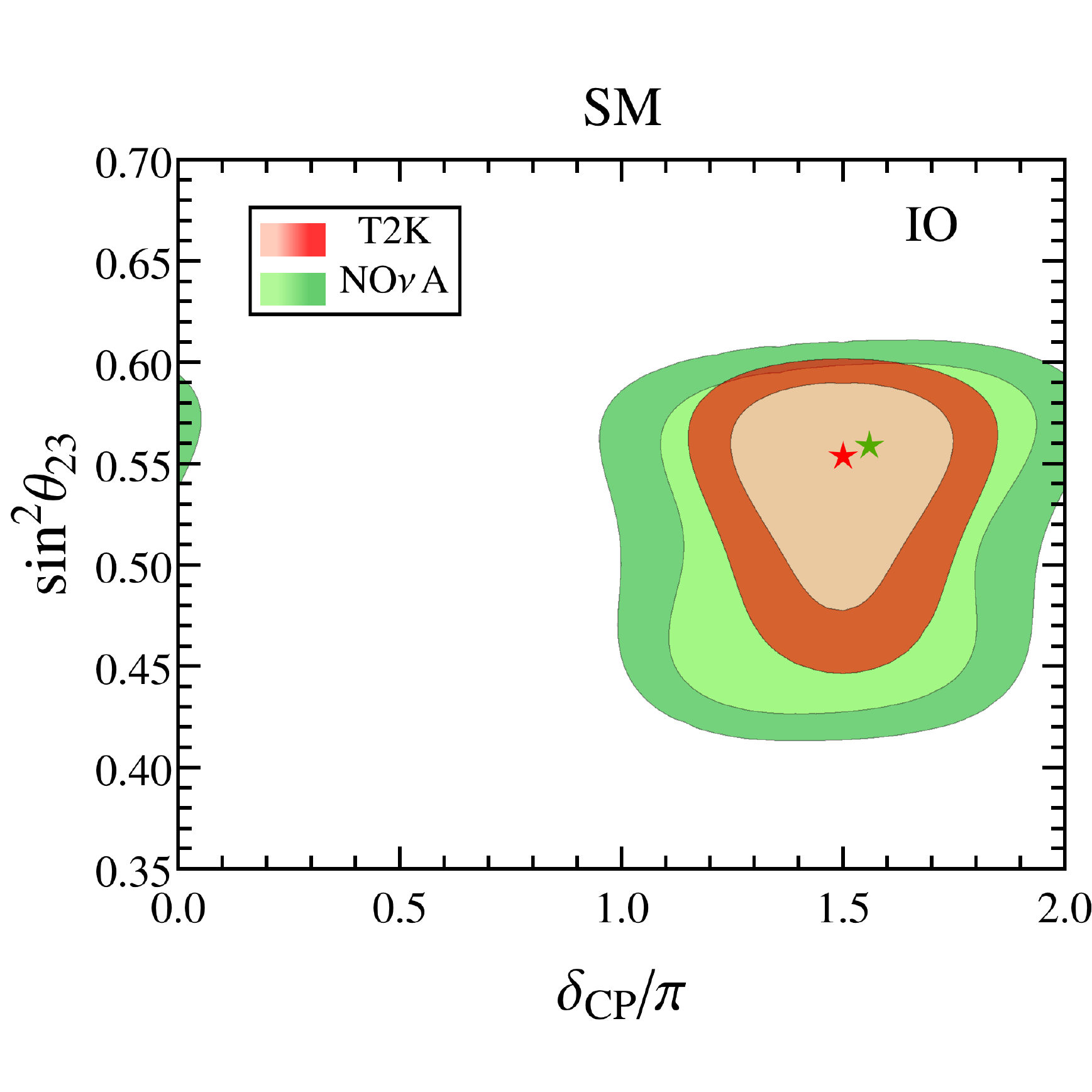}
\includegraphics[height=5.87cm,width=5.87cm]{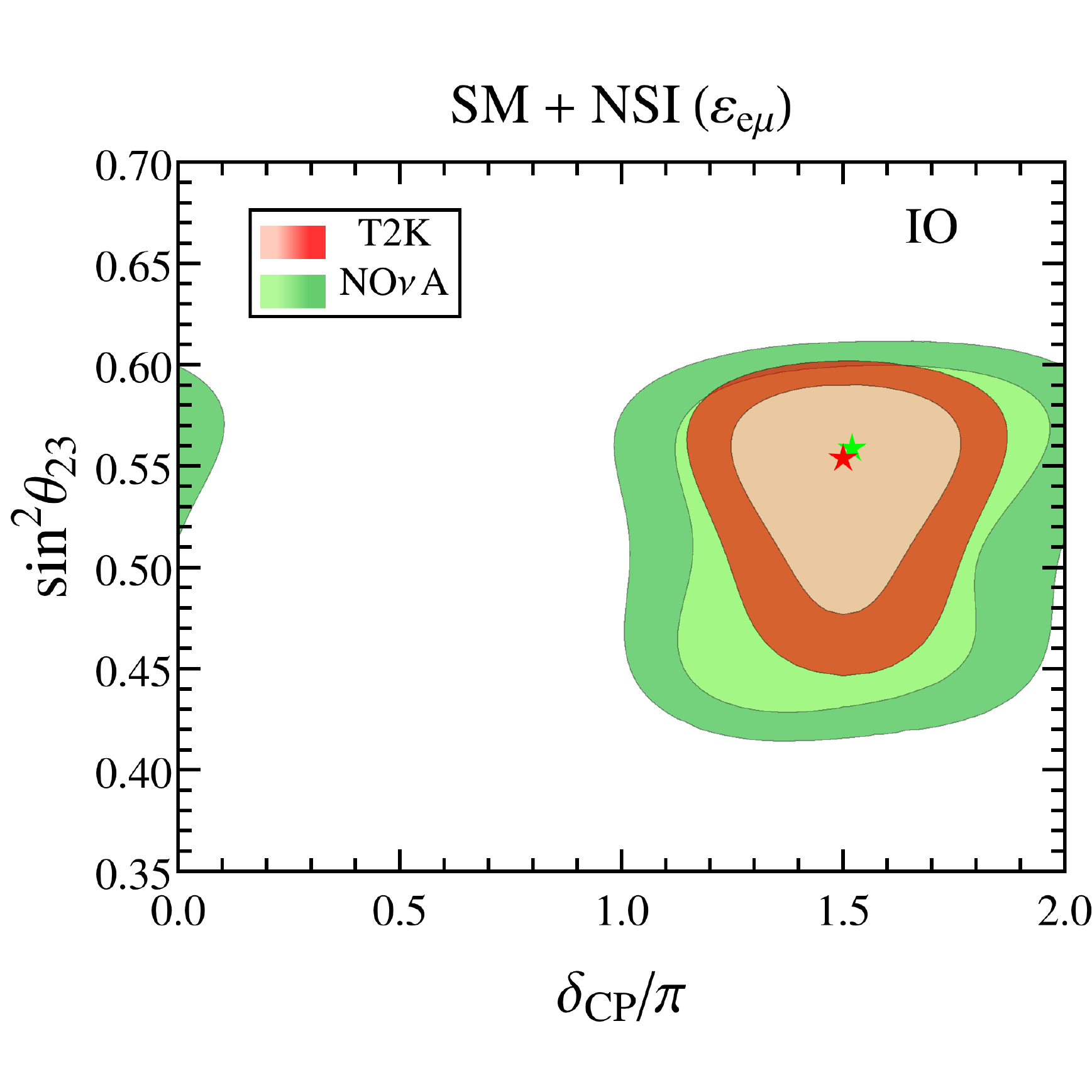}
\includegraphics[height=5.87cm,width=5.87cm]{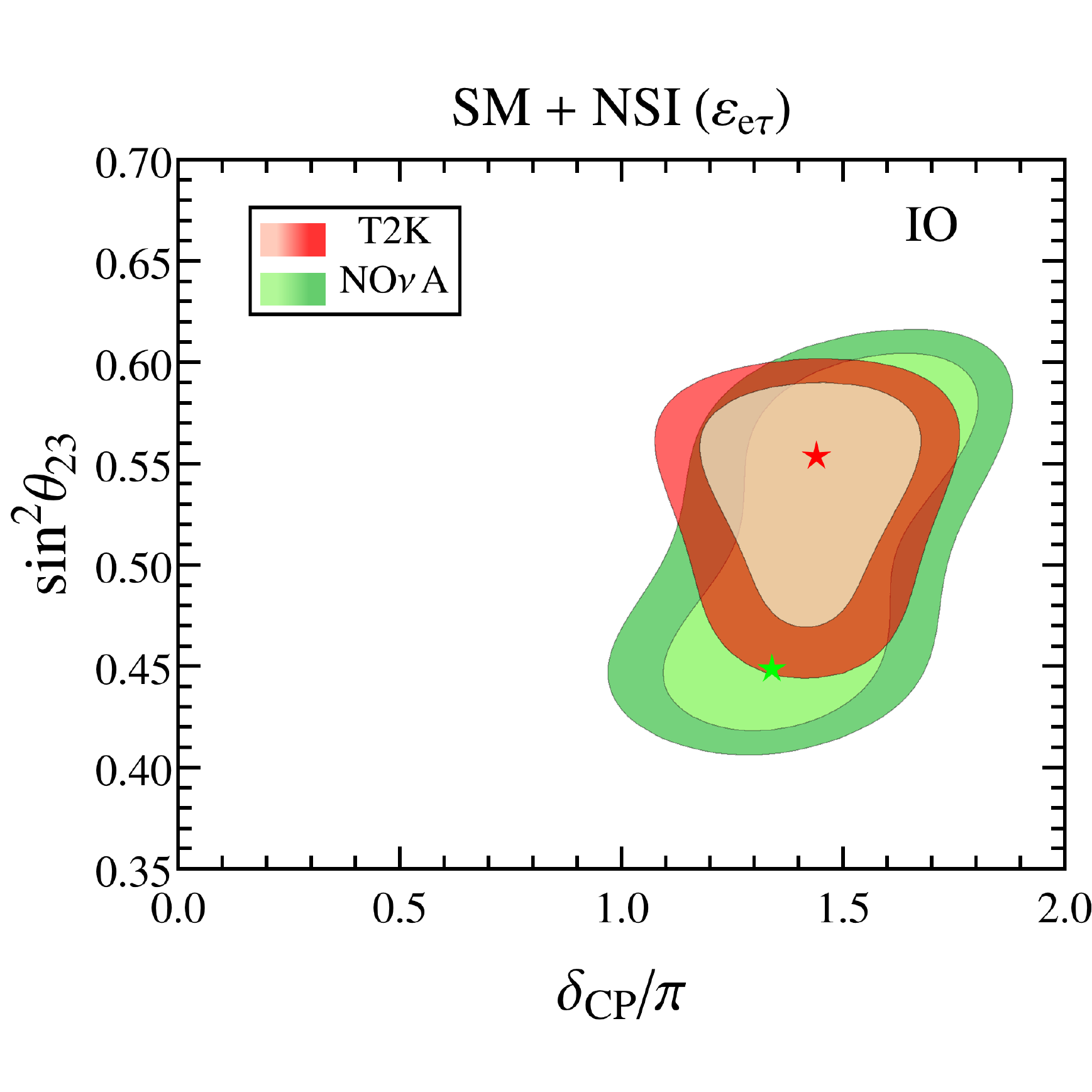}
\vspace*{-0.5cm}
\caption{Allowed regions determined separately by T2K and NO$\nu$A for IO in the SM case
(left panel) and with NSI in the $e-\mu$ sector (middle panel) and in the $e-\tau$ sector
 (right panel). In the middle panel we have taken 
the NSI parameters at their best fit values of T2K + NO$\nu$A ($|\varepsilon_{e\mu}| = 0.02, \phi_{e\mu} = 0.96\pi$).
Similarly, in the right panel we have taken $|\varepsilon_{e\tau}| = 0.15, \phi_{e\tau} = 1.58\pi$.
The contours are drawn at the 68$\%$ and 90$\%$ C.L. for 2 d.o.f. The comparison of the middle and right panels
with the left one neatly shows that, in the IO case, there is no improvement when adding the NSI since
the values of $\delta_{\rm CP}$ identified by T2K and NO$\nu$A are in excellent agreement in the
SM case.}
\label{fig_tension_IO}
\end{figure*}

Figure~\ref{fig_tension_IO} is analogous to Fig.~3, but refers to the IO case. As already evident 
from Figs.~1 and 2, in the IO case there is basically no preference for non-zero NSI.  
This happens because, differently from the NO case, the SM is able to fit well the data of T2K and NO$\nu$A  
with the same value of the CP phase $\delta_{\rm CP}$ around 1.5$\pi$. This is evident in the left panel of Fig.~\ref{fig_tension_IO}.
From the middle and right panels of the same figure, one sees that adding a NSI does not 
help to improve the fit, since the agreement of the two experiments is already very good in the SM case.

\begin{figure*}[h!]
\vspace*{-0.0cm}
\hspace*{-0.2cm}
\includegraphics[height=4.9cm,width=4.9cm]{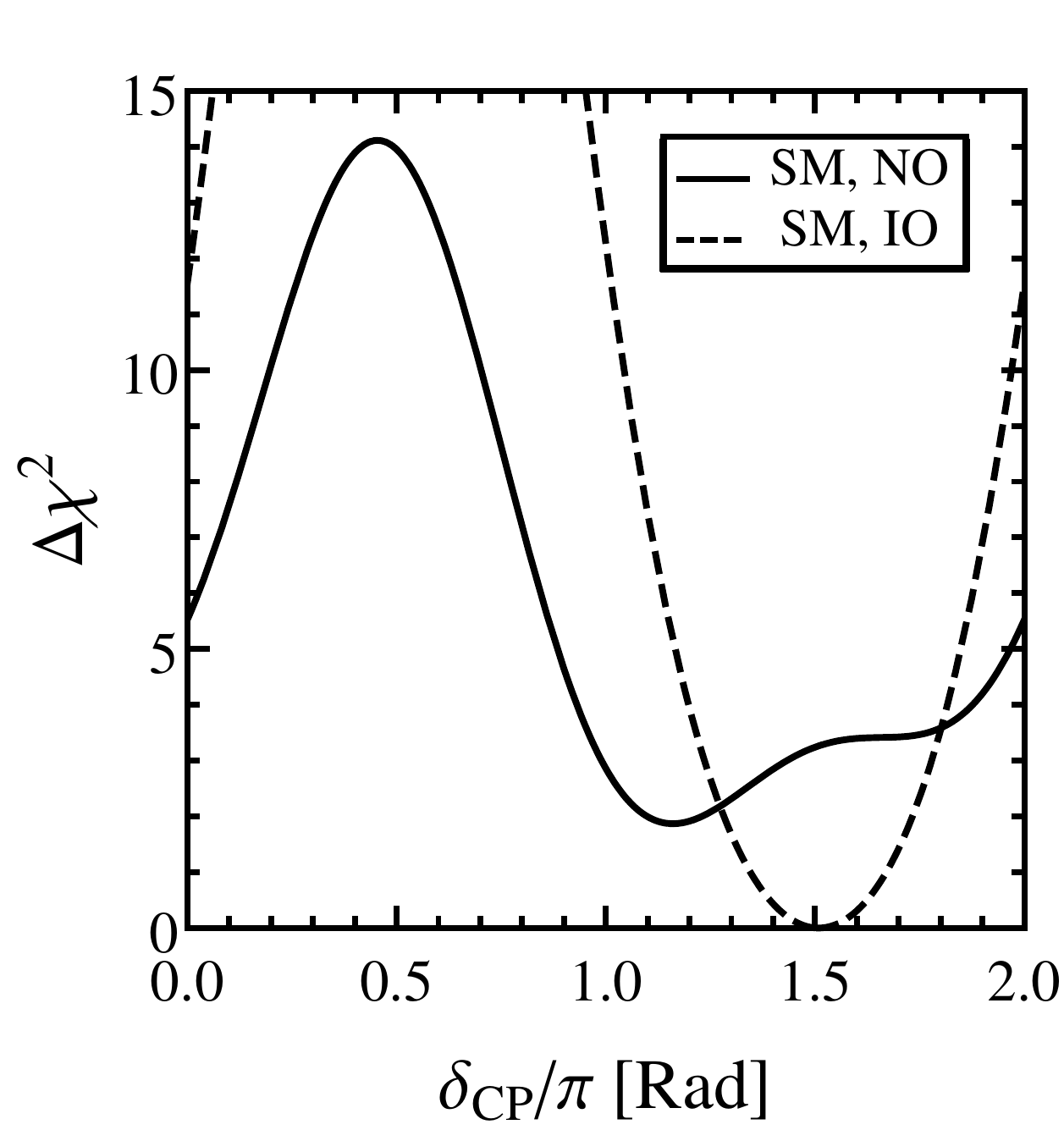}
\includegraphics[height=4.9cm,width=4.9cm]{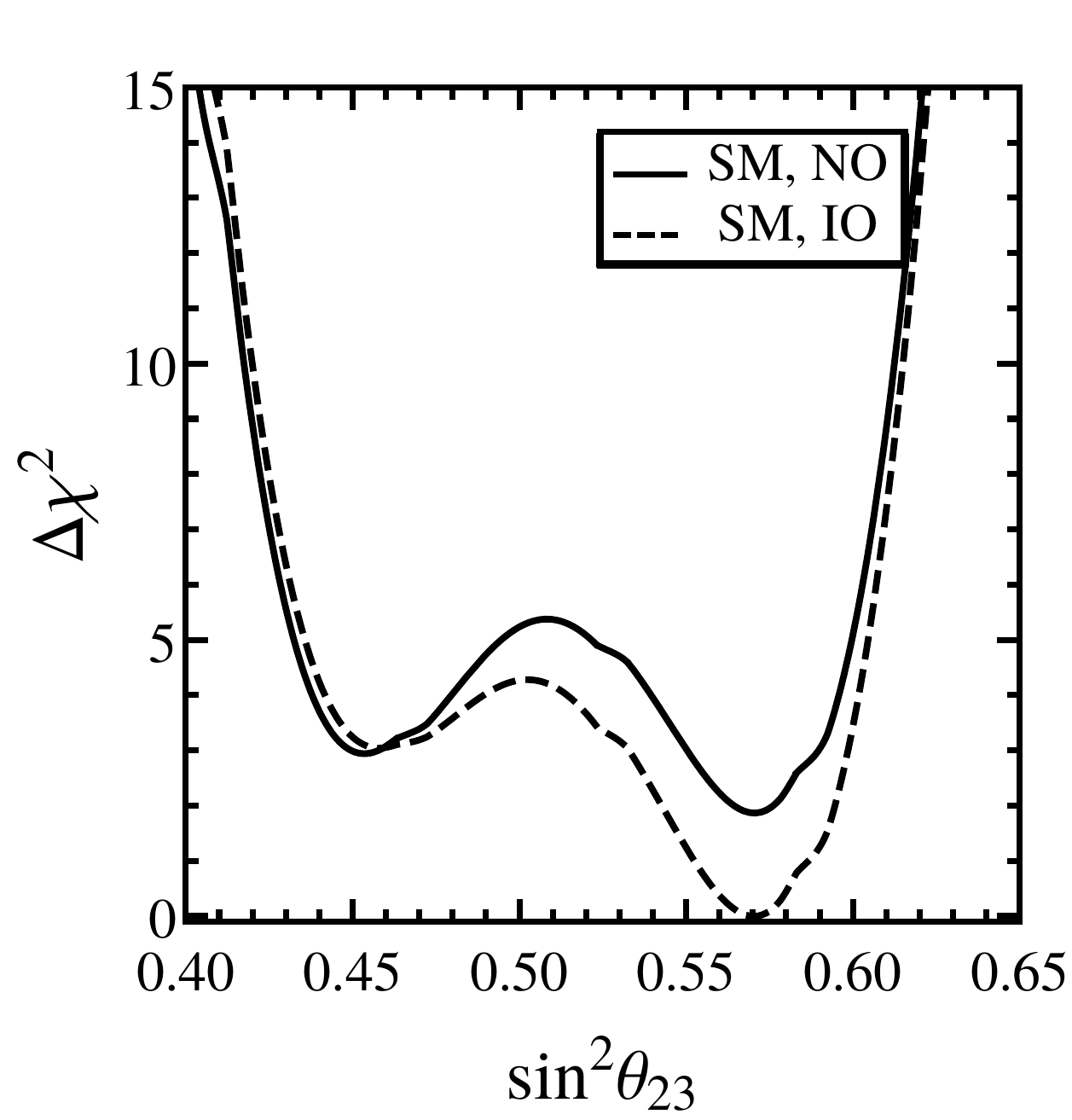}
\includegraphics[height=4.9cm,width=4.9cm]{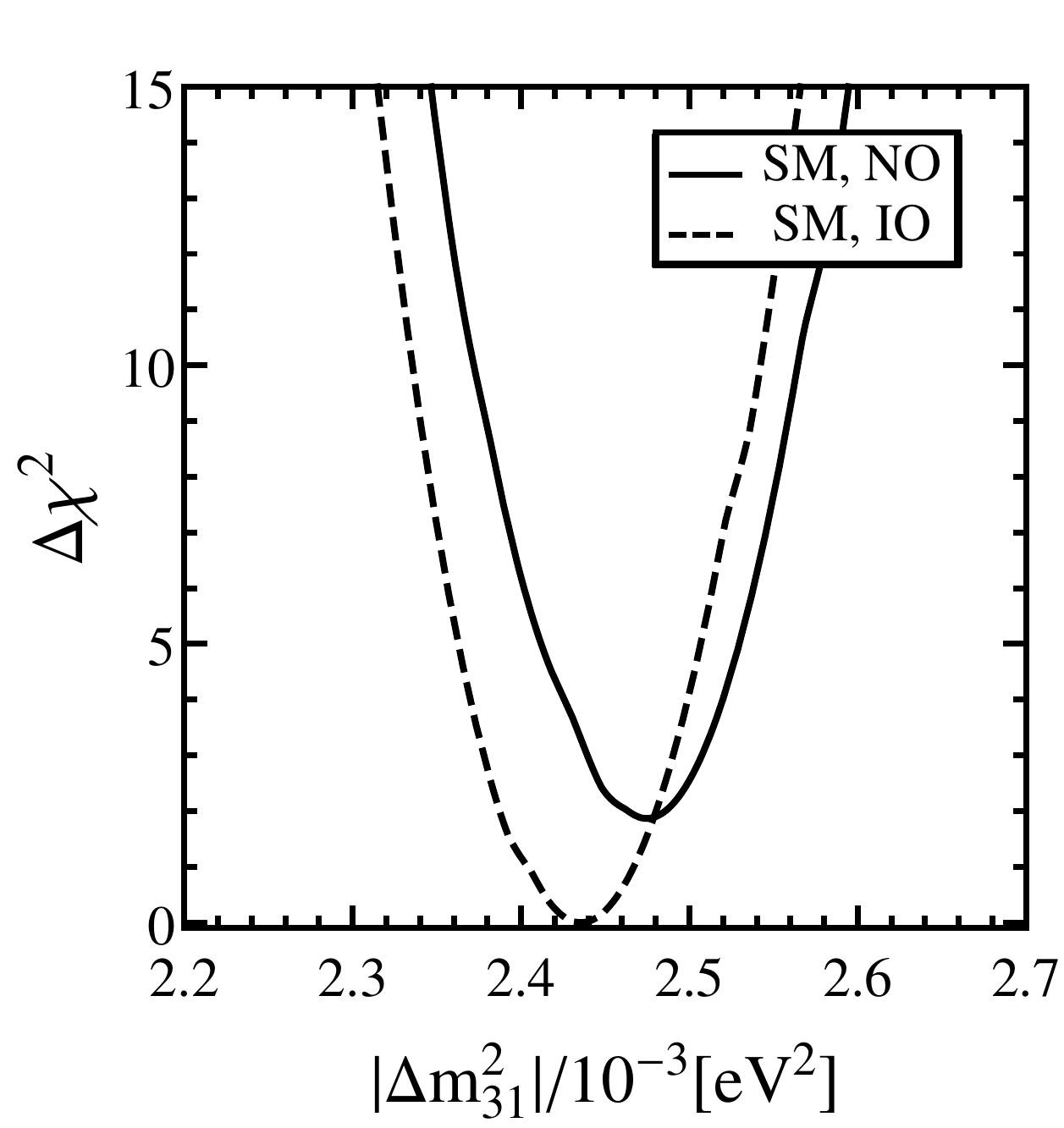}\\
\includegraphics[height=4.9cm,width=4.9cm]{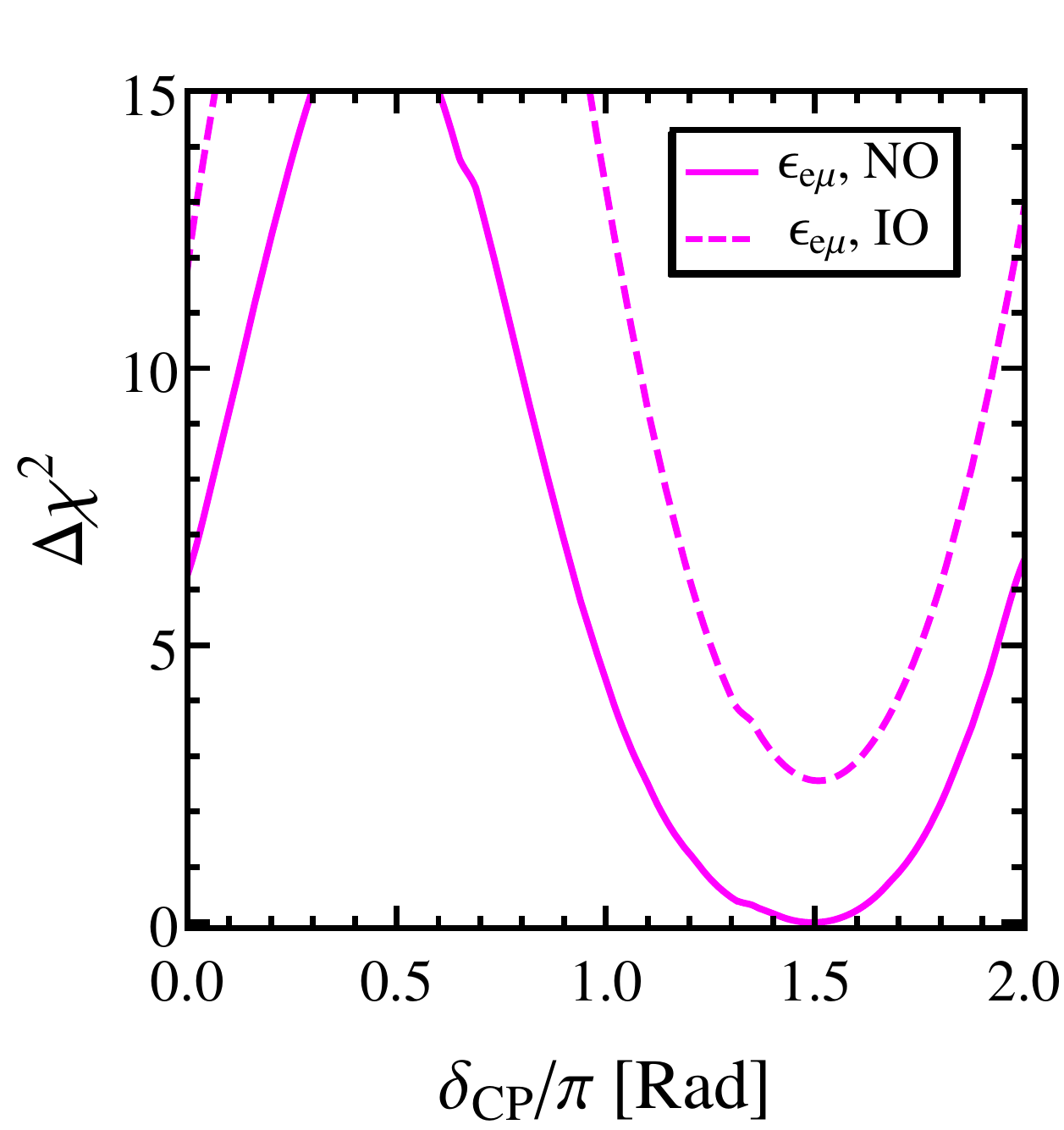}
\includegraphics[height=4.9cm,width=4.9cm]{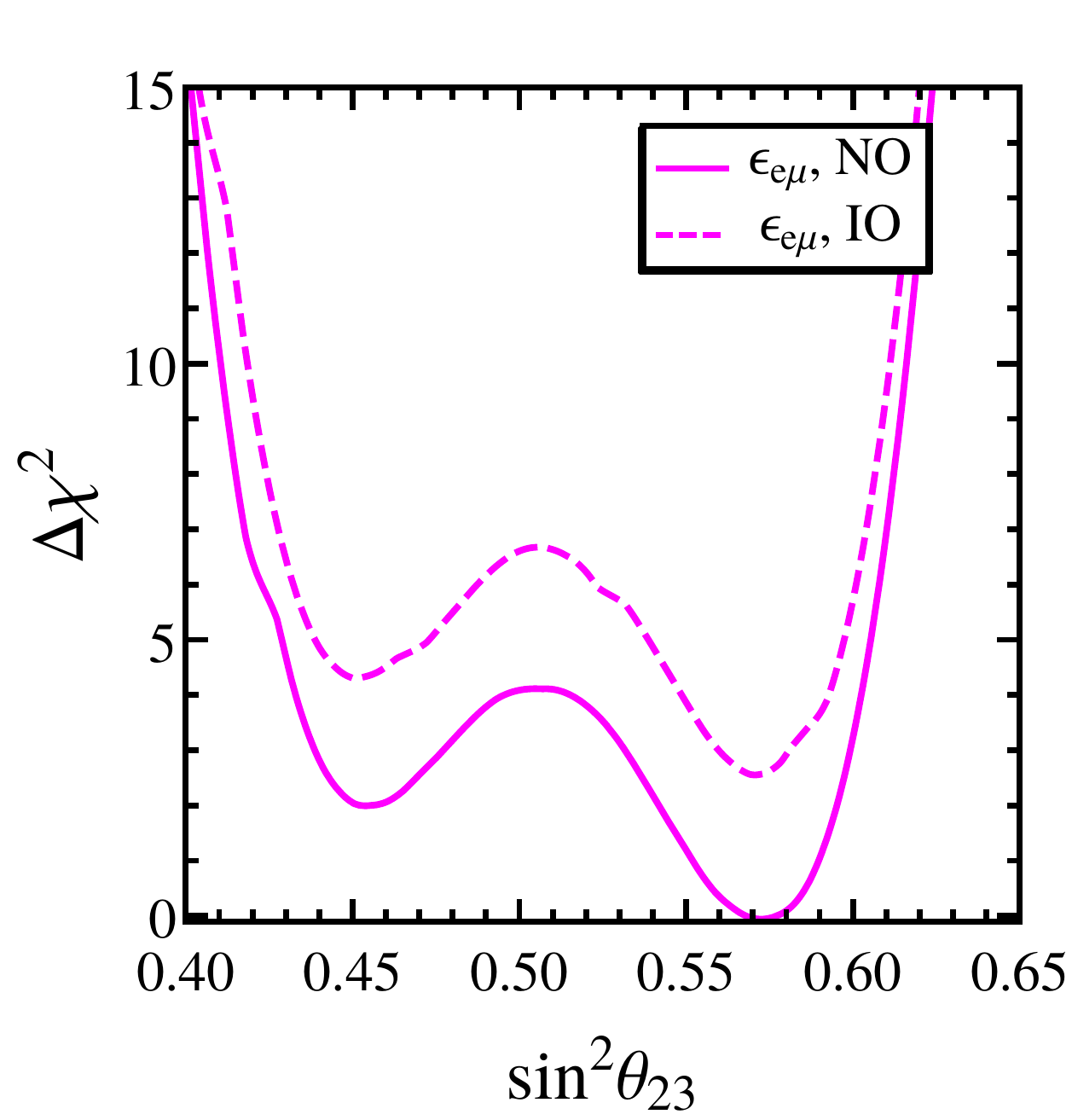}
\includegraphics[height=4.9cm,width=4.9cm]{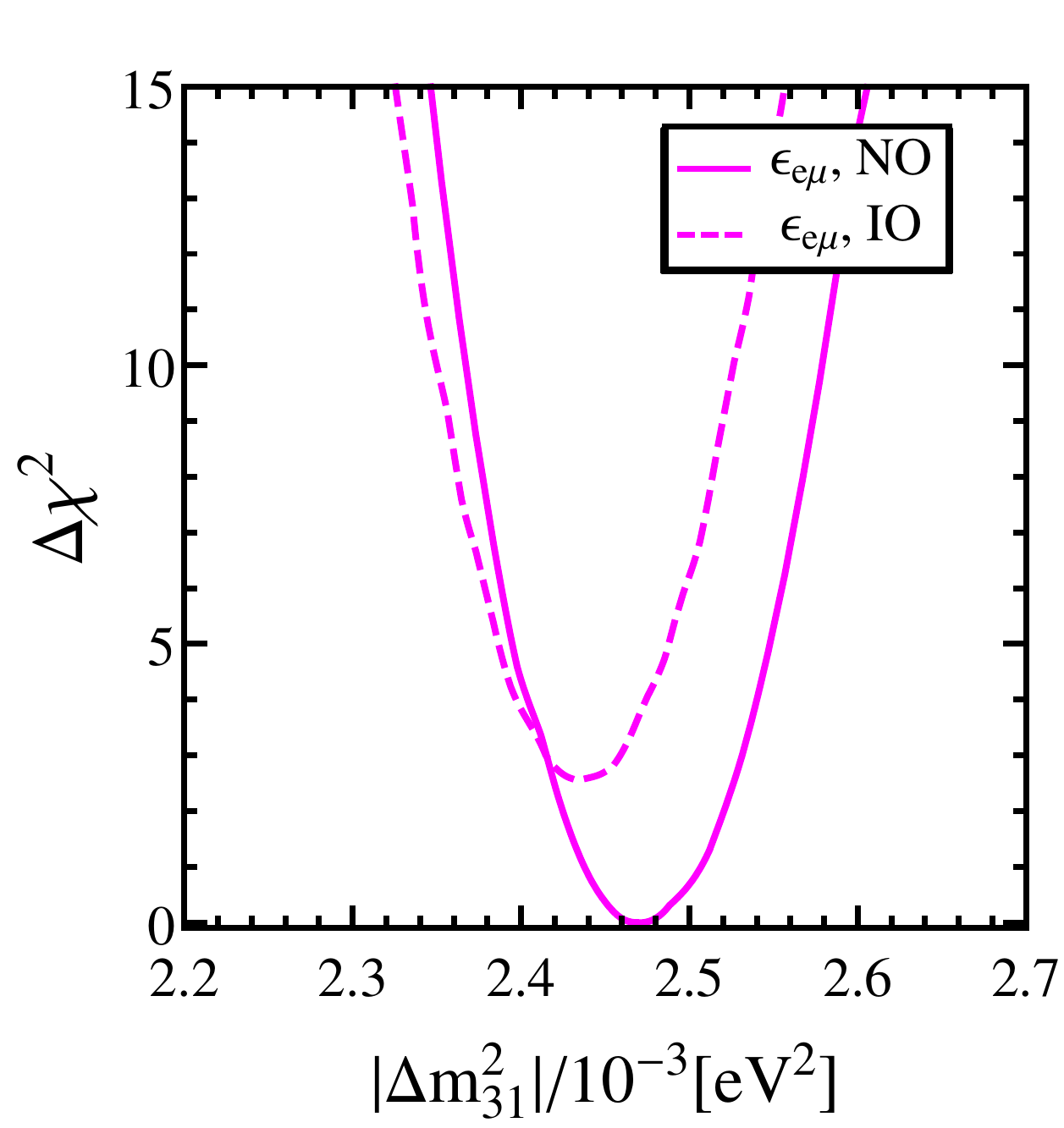}\\
\includegraphics[height=4.9cm,width=4.9cm]{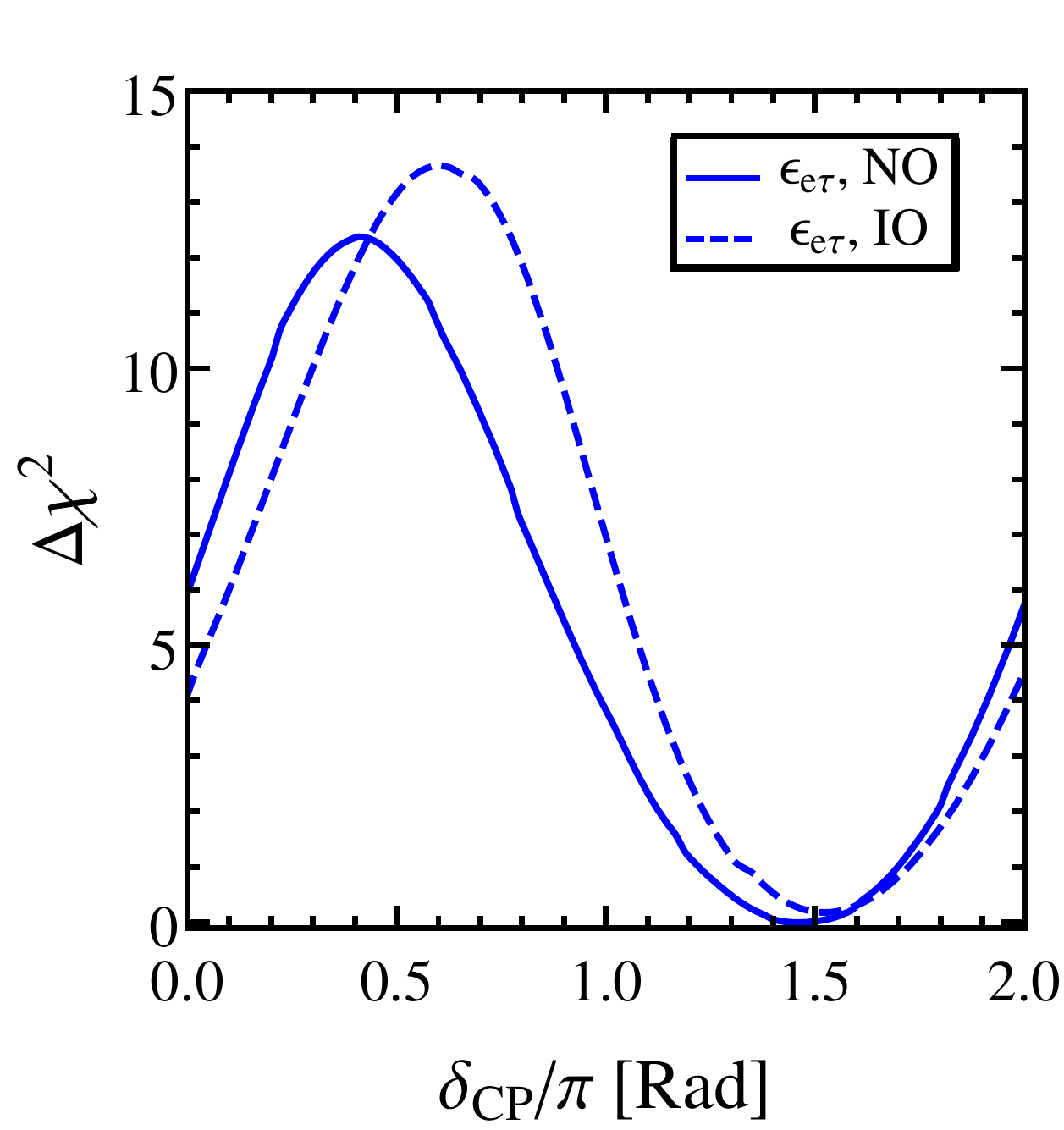}
\includegraphics[height=4.9cm,width=4.9cm]{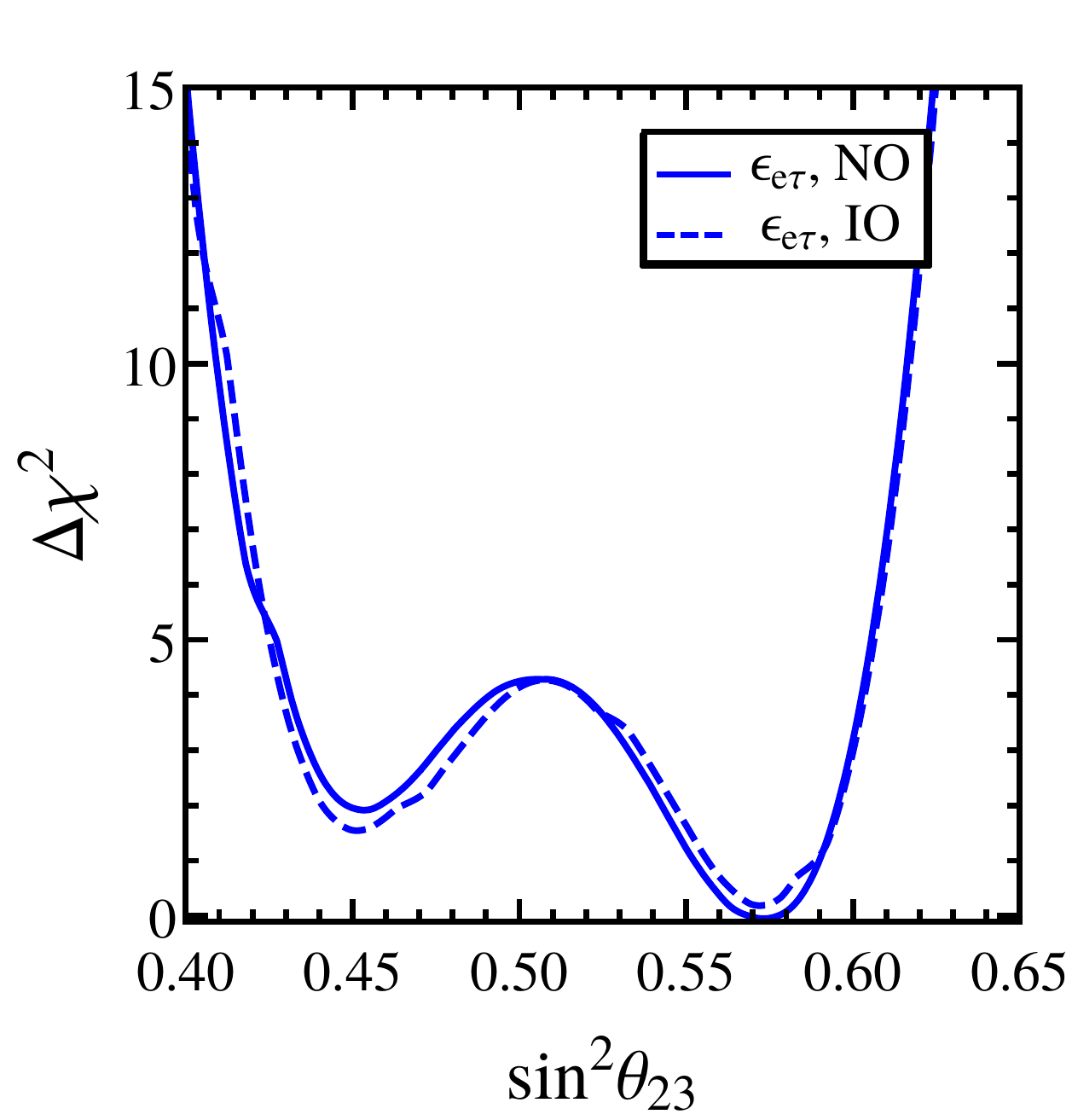}
\includegraphics[height=4.9cm,width=4.9cm]{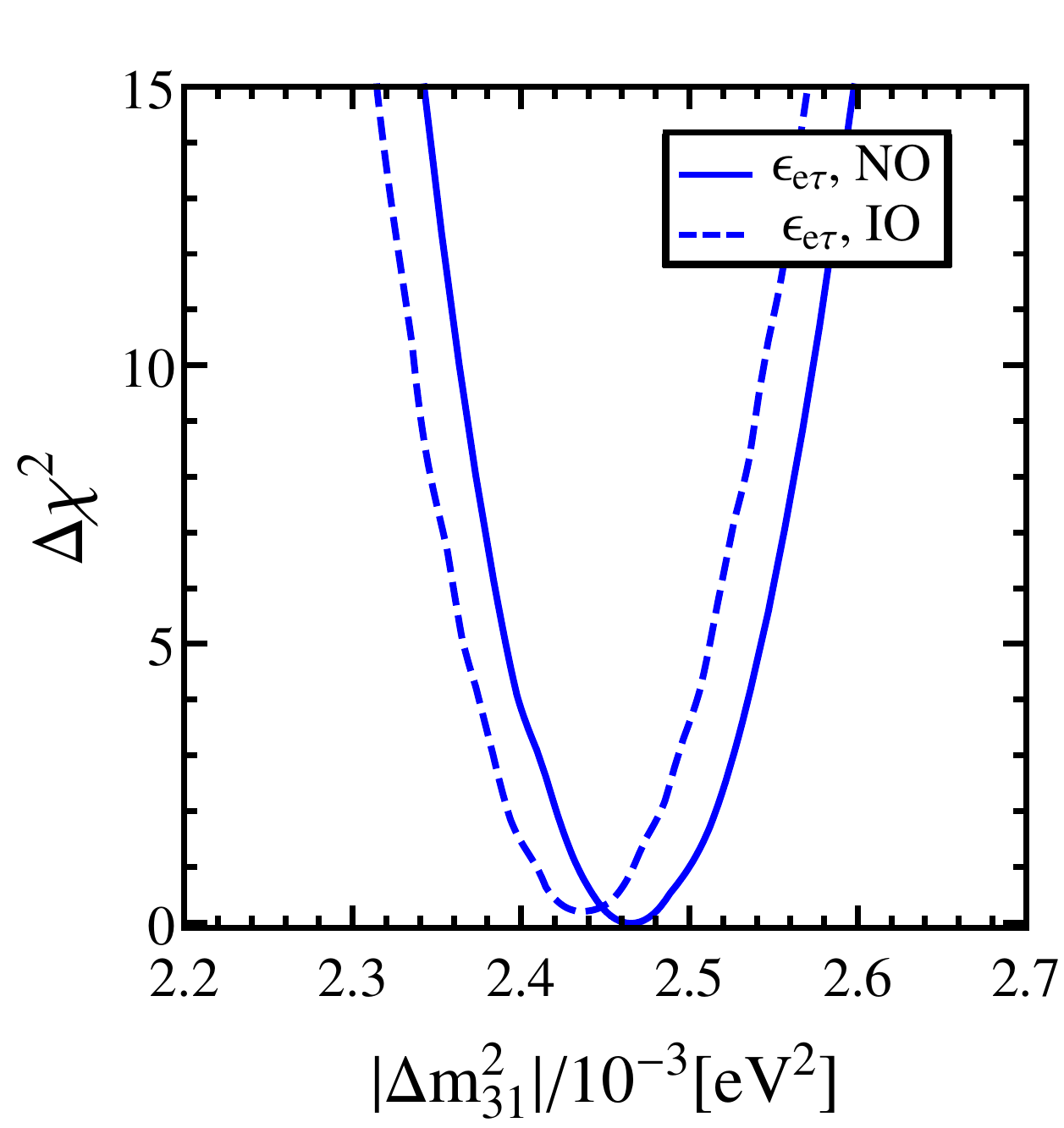}
\vspace*{-0.0cm}
\caption{One-dimensional projections of the standard parameters determined by the combination of T2K and NO$\nu$A for NO (continuous curves) and IO (dashed curves).}
\label{fig:regions_1D}
\end{figure*}

Fig.~\ref{fig:regions_1D} reports the one-dimensional projections on the standard 
oscillation parameters $\delta_{\mathrm {CP}}$, $\theta_{23}$ and $|\Delta m^2_{31}|$
from the combination of NO$\nu$A and T2K attained by expanding
the $\chi^2$ around the minimum value obtained when the SM, SM+NSI ($\varepsilon_{e\mu}$)
and the SM+NSI ($\varepsilon_{e\tau}$) hypotheses are accepted as true.
 The upper, middle and lower panels refer respectively
to the SM case, the SM+NSI in the $e-\mu$ sector, and the SM+NSI in the  $e-\tau$ sector.
The continuous (dashed) curves correspond to NO (IO). The left upper panel evidences in the NO
case an oscillating behavior of the CP-phase $\delta_{\mathrm {CP}}$. This is the result of the
discrepant values preferred by the two experiments. In the presence of NSI (middle left and lower left panels)
 this oscillating behavior disappears as both experiments point towards the same common value
$\delta_{\mathrm {CP}} \sim3\pi/2$. Concerning the neutrino mass ordering, we note that in the SM case,
as found in other recent analyses~\cite{Kelly:2020fkv,Esteban:2020cvm,Denton:2020uda},
 there is a slight preference for IO ($\chi^2_{\mathrm{NO}} - \chi^2_{\mathrm{IO}} = 1.87$). 
In the presence of NSI there is a moderate preference for NO for NSI in the $e-\mu$ sector ($\chi^2_{\mathrm{NO}} - \chi^2_{\mathrm{IO}} = -2.56$), while no ordering is favored for NSI in the $e-\tau$ sector ($\chi^2_{\mathrm{NO}} - \chi^2_{\mathrm{IO}} = -0.21$).
One can also compare the preference of NO in presence of NSI with respect to the IO in the SM case obtaining
$\chi^2_{\rm e\mu, NO} - \chi^2_{\rm SM, IO} = -2.63$ and $\chi^2_{\rm e\tau, NO} - \chi^2_{\rm SM, IO} = -1.21$.
In both cases, a slight preference for NO appears. This means that NO with NSI is more likely explanation than IO 
of NO$\nu$A-T2K discrepancy either with or without NSI.  Minor differences appear in the estimate of $\theta_{23}$  among the three
cases. In all scenarios, non-maximal $\theta_{23}$ mixing in the second octant is slightly favored.

\end{document}